%% file: Termination.tex
\documentclass{llncs}
\usepackage{graphicx}			
\input{MacrosGeneral}

\input{Macros}
\input{MacrosMarkup}
\Bleck 

\begin{document}
\input{MacrosLabels} 
\ifBleck\else\ShowLabels\fi
\MakeMarkupsCalledtrue
\MakeMarkups[Carroll]{C}{\Colour{blue}}
\MakeMarkups[Annabelle]{A}{\Colour{red}}

\title{A new rule for almost-certain termination \\ of probabilistic- and demonic programs}
\author{Annabelle McIver\inst{1} and Carroll Morgan\inst{2}}
\institute{Macquarie University, Australia \and University of New South Wales, and Data61, Australia.}
\maketitle
\thispagestyle{plain} 

\begin{abstract}
Extending our own and others' earlier approaches to reasoning about termination of probabilistic programs, we propose and prove a new rule for termination with probability one, also known as ``almost-certain termination''. The rule uses  both (non-strict) super martingales and guarantees of progress, together, and it seems to cover significant cases that earlier methods do not. In particular, it suffices for termination of the unbounded symmetric random walk in both one- and two dimensions: for the first, we give a proof; for the second, we use a theorem of Foster to argue that a proof exists.

Non-determinism (i.e.\ demonic choice) is supported; but we do currently restrict to discrete distributions.
\footnote{This revises an earlier version \cite{McIver:2016aa} by correcting typographical errors and adding an extra section \Sec{s1031} on historical background.}
\Cf{Our programming logic (recall) assumes \emph{bounded} nondeterminism. We will have to address that in the ``full version''.}
\end{abstract}

\Section{Introduction}

This paper concerns proof of almost-sure termination for probabilistic- and demonic programs, ones that move from one state to another by first choosing a (discrete) distribution demonically from a set of distributions, and then choosing a new state probabilistically according to that distribution.

Thus we view a program abstractly as a (probabilistic/demonic) transition system; and we are interested in proving the eventual reachability with probability one of a given set of target states (when it is indeed the case). Our strategic aim is to express our techniques in a form that  can be applied to probabilistic program code \emph{in situ}, i.e.\ without the need to construct the programs' underlying transition systems explicitly (although of course we rely on their existence). That is, we seek (and find) proof-rules that require no more than local reasoning in the source code.

In probabilistic programming over a finite state-space $S$, say, a typical rule is one that generalises the ``variant rule'' for standard (non-probabilistic but still demonic) programs: to each state is assigned a non-negative integer, a \emph{variant} bounded above and below, with all states inside some target set $S_0{\subseteq}S$ assigned variant $0$. One then shows by local reasoning, typically over the source-code of a loop body,
\begin{Equation}\label{e1511}
\begin{minipage}{0.7\linewidth}
that from each state in $S_\ast=S{-}S_0$ there is a \emph{non-zero probability}
of transiting to a (different) state with strictly smaller (integer) variant.\quad\footnotemark
\end{minipage}
\end{Equation}%
If that can be shown, then the rule guarantees that almost all  paths in the transition system eventually lead to  $S_0$, where ``almost all'' means that the paths not included have probability zero even if taken all together.
\footnotetext{This is the probabilstic generalisation: in the traditional, non-probabilistic rule the decrease must be certain. On the other hand, in the traditional case the variant need not be bounded above. In both cases, it must be bounded below.}
This probabilistic rule's  soundness  follows  from an appeal to a zero-one law  \cite{Hart:83,Morgan:96b, McIver:05a} which roughly says:

\begin{quote}
If there is some $\varepsilon {>} 0$ such that the  probability of eventually reaching a target set of states is \emph{everywhere} at least $\varepsilon$, then that probability is one. 
 \end{quote}

For infinite-state systems however, although such zero-one laws are still valid, their $\varepsilon$-conditions are not so easily met by local reasoning.  In particular the actual values of the probabilities attached to the transitions, which in fact are irrelevant in finite-state transition systems \cite{Morgan:03,McIver:03}, now can make the difference between  almost-certain termination or not. A typical response to this issue is to replace ``non-zero probability'' in \Eqn{e1511} above with ``probability bounded away from zero''. And that bound can depend intricately on the transitions' actual probabilities.

That challenge notwithstanding, recent important work \cite{Chakarov:2013,Fioriti:2015,Chakarov:2016aa} has shown how local reasoning with super-martingales can be applied to solve the termination problem in a wide class of infinite-state probabilistic programs.

In this paper we combine those successes with some of our own earlier work, showing in this paper how to use super-martingale reasoning together with a progress rule to reason about to an important class of transition systems whose termination seems to be beyond the state-of-the-art, for source-level reasoning at least.  Our key insight is the observation that the combination of a super-martingale with a local \emph{but parametrised} progress-condition (in a sense we explain below) implies the conditions of the zero-one rule. 

Our specific contributions are:
\begin{enumerate}
\item A new rule that generalises a number of currently known rules (including our own) for establishing almost-certain termination;
\item A demonstration of a general zero-one proof technique which can be applied in arbitrary infinite state systems;
\item A thorough analysis of the applicability of the new rule together with a suite of representative examples; and
\item A limited survey of some pre- computer-science mathematical results that contribute to this endeavour \cite{Kendall:1951aa,Foster:1952aa,Blackwell:55}
\end{enumerate}

Finally we note that our strategic goal, to translate this and other rules  to ones that can be applied directly to program code, lies in the seminal work of Kozen \cite{Kozen:85} for probabilistic semantics, later generalised \C{by us} to include demonic non-determinism and abstract transition systems \cite{Morgan:96d,McIver:05a} and even more recently expanded to include explicit Markov-chain models \cite{GretzKM14}.

\Section{Informal description of the new rule for termination}
\Subsection{Setting of the new rule, and its purpose}
Let $S$ be a state space, possibly infinite, and let $S$ be divided into two disjoint subsets: one is $S_0$, the states where termination is deemed to have occurred; and the other is $S_\ast$ for the rest. A transition function $T$ is given taking any state $s$ in $S_\ast$ to a set of discrete distributions on all of $S$; and a transition from some $s$ in $S_\ast$ occurs by first selecting arbitrarily a distribution $\delta$ from $T(s)$ and then choosing probabilistically a next-state $s'$ according to the probabilities given by $\delta$. (We discuss this treatment of demonic nondeterminism more thoroughly in \Sec{s0910a} below.)

Our purpose is to give a method for proving that from any $s$ in $S_\ast$, repeated transitions according to $T$ will reach $S_0$ with probability one eventually. That property is conventionally called \emph{almost-certain termination}. For brevity, from now on we will write \AC\ for ``almost certain(ly)'' and \ACT\ for ``almost-certain(ly) terminat(ion/ing)''.


\Subsection{Informal description of the rule}\label{s1203}
By analogy with existing approaches to proof of termination, we base our technique on a ``variant'' function over the states and require it to have certain properties. Informally described, they are as follows:
\begin{description}
\item[Define a \textit{variant} ---]
Non-negative variant function $V$ from (all of) $S$ into the non-negative reals is such that $V$ is 0 on all of $S_0$ and $V$ is strictly positive on $S_\ast$. It can be unbounded (above), but not infinite. Note that $V$ need not be integer-valued.


\item[Impose a \SMart\ property ---] Variant $V$ is a super-martingale wrt.\ transitions $T$, i.e.\ for any $s$ in $S_\ast$ and any distribution $\delta$ in $T(s)$, the expected value of $V$ on $\delta$, i.e.\ on the states reached in one $\delta$-mediated transition of $T$ from $s$, is no more than the value $V(s)$ that $V$ had at $s$ itself. That is,
\[
 \text{For all $s$ in $S_\ast$ and $\delta$ in $T(s)$ we have}\hspace{5ex} \Exp_\delta V \leq V(s)\quad,
\]
where we write $\Exp_\delta V$ for the expected value, over discrete distribution $\delta$ on $S$, of real-valued function $V$ on $S$.

Note that we do not require a strict decrease of the expected value, and although $V$ is defined on $S_0$, we do not require that $T$ be defined there.
\footnote{Although in general there is a question of definedness of $\Exp_\delta V$ when $\delta$ has infinite support and $V$ is unbounded, that does not arise here.}

\item[Impose a \Progress\ property ---] The transitions $T$ makes progress towards $S_0$. We require two fixed strictly positive functions $p$ (for ``probability'') and $d$ (for ``decrease''), defined for all positive reals, such that in a state $s$ of $S_\ast$ with $V(s)$ equal to some $v$, any transition $\delta$ in $T(s)$ is guaranteed with probability at least $p(v)$ to decrease the variant by at least $d(v)$. Furthermore $d(v)$ and and $p(v)$ must be non-increasing as $v$ itself increases. That is,

\smallskip
\begin{quote}
 There are fixed functions $p,d$ on positive reals $v$, with $0{<}p(v){\leq}1$ and $0{<}d(v)$, such that whenever  $v{=}V(s)$ for some $s$ in $S_\ast$, and $\delta$ in $T(s)$, we have
 \[
  \delta_{\{s'\mid V(s'){\leq}v{-}d(v)\}}\Wide{\geq}p(v)~,
  \hspace{6ex}\parbox[t]{25ex}{\small where for $S'{\subseteq}S$\\ we write $\delta_{S'}$ for $\sum_{s'\In S'} \delta_{s'}$~,}
 \]
\end{quote}
and for any $0{<}v{<}v'$ we have $p(v'){\leq}p(v)$ and $d(v'){\leq}d(v)$.
\end{description}
Note that $p,d$ in \Progress\ are functions of the \emph{variant}, defined over all positive reals, and that even for $v$ not in the $V$-image of $S$, still the non-increasing conditions for $p(v),d(v)$ must be satisfied. (See \Sec{s1905}'s ``What happens when $V$ is bounded''.)

The rule is proved in \Sec{s0839} below.

\Subsection{Discussion and comparison}\label{ss1958}

Our main innovation in \Sec{s1203} is, in our \Progress\ condition, to impose the usual ``bounded away from zero'' criterion not on $S_\ast$ as a whole but instead only on successively larger subsets of it. That is, we apply it with respect to certain functions $p$ and $d$, and the effect of their non-increasing criteria is to ensure that, as the subsets $\{s{\In}S_\ast\mid V(s){\leq}v\}$ of $S_\ast$ grow larger, the progress conditions imposed on them grow weaker but never decrease to ``none''. This avoids the treacherous Zeno-effects that can occur when some progress is always made but only with ever-smaller steps: the $V$-decrease condition (``\,as far as $d$ with probability at least $p$\,'') can only be strengthened as $V(s)$ moves towards 0. But it also avoids the need to set a uniform $\varepsilon$-progress condition for all of $S_\ast$.

Although the generality of $p,d$ might seem complicated, in fact in many special cases it is very simple. One such is the ``distance from 0'' variant on the one-dimensional symmetric random walk, where $p,d$ can be constant functions: we take $S$ to be the integers, both positive and negative, with $S_0{=}\{0\}$ and $V(s){=}\Abs{s}$ and we define $p,d$ to be everywhere $\NF{1}{2},1$ respectively --- with probability at least $\NF{1}{2}$ the variant decreases by at least 1. That is all that's needed to establish \ACT\ for the symmetric random walk (\Sec{s0910}).
\footnote{This simplicity shows that the difficulty of finding an \ACT\ rule lies in part in making sure it does not allow too much: what prevents our rule's proving that a biased random walk is \ACT? See \Sec{s0904a}.}

\Subsection{Other approaches}
In \textbf{our own, earlier probabilistic-variant rule} \cite[Sec.\,6]{Morgan:96b},\cite[Sec.\,2.7]{McIver:05a}, we effectively made $p,d$ constants, imposed no \SMart\ condition but instead bounded $V$ above over $S_\ast$, making it not sufficient for the random walk. \textbf{Later however} we did prove random walk to be \ACT\ using a rule more like the current one \cite[Sec. 3.3]{McIver:05a}.

\medskip
\textbf{Chakarov and Sankaranarayanan} \cite{Chakarov:2013} consider the use of martingales for the analysis of infinite-state probabilistic programs, and \textbf{Chakarov} has done more extensive work \cite{Chakarov:2016aa}.

In \cite{Chakarov:2013} it's shown that a \emph{ranking super-martingale} implies \ACT, and a key property of their definition for ranking super-martingale is that there is some constant $\varepsilon {>}0$ such that the average decrease of the super-martingale is everywhere (except for the termination states) at least $\varepsilon$. Their program model is  assumed to operate over discrete state spaces, without nondeterminism.

That work is an important step towards applying results from probability theory to the verification of infinite-state probabilistic programs.

\medskip
\textbf{Fioriti and Hermanns} \cite{Fioriti:2015} also use ranking super-martingales, with results that provide a significant extension to Chakarov and Sankaranarayanan's work \cite{Chakarov:2013}. Their program model includes both non-determinism and continuous probability distributions over transitions. They also show completeness for the class of programs whose expected time to termination is finite.  That excludes the random walk however; but they do demonstrate by example that the method can still apply to some systems which do not have finite termination time.

\medskip
More recently still, \textbf{Chatterjee, Novotn\'y and {\v Z}ikeli\'c} \cite{Chatterjee:16} study techniques for proving that programs terminate with some probability (not necessarily one). Their innovation is to introduce the concept of ``repulsing super-martingales'' --- these are also super-martingales  with values that decrease outside of some defined set. Repulsing super-martingales can be used to show lower bounds on termination probabilities, and as certificates to refute almost-sure termination and finite expected times to termination. (See also \Sec{s1207}, \Sec{ss0829}.)

\medskip
There are a number of other works that demonstrate tool support based on the above and similar techniques.  All the authors above \cite{Chakarov:2013,Fioriti:2015,Chatterjee:16} have developed and implemented algorithms to support verification based on super-martingales.  \textbf{Esparza, Gaiser and Kiefer} \cite{Easparza:12} develop algorithmic support for \ACT\ of ``weakly finite'' programs, where a program is \emph{weakly finite} if the set of states reachable from any initial state is finite.  \textbf{Kaminski et al.}\ \cite{Kaminski:16} have studied the analysis of expected termination times of infinite state systems using probabilistic invariant-style reasoning, with some applications to \ACT. In even earlier work \textbf{Celiku and McIver} \cite{Celiku:05} explore the mechanisation of upper bounds on expected termination times, taking probabilistic weakest pre-conditions \cite{McIver:05a} for their model of probability and non-determinism.

\Section{Our treatment of demonic nondeterminism}\label{s0910a}
Before proving \Sec{s1203}, we explain our treatment of demonic- and probabilistic choice together.

Our transition function $T$ is of type $S_\ast{\Fun}\Pow\Dist S$, where $\Pow$ is the powerset constructor and $\Dist$ is the discrete-distribution constructor: thus for state $s$ in $S_\ast$, its possible transitions $T(s)$ comprise a set ($\Pow$) of discrete distributions ($\Dist$) of states ($S$). It simultaneously extends (1) the conventional model $S{\Fun}\Pow S$ of demonic (non-probabilistic) programs and e.g.\ (2) Kozen's model $S{\Fun}\Dist S$ \cite{Kozen:85} and later Plotkin and Jones' model \cite{Jones:89} of probabilistic (non-demonic, i.e.\ deterministic) programs. For (1) the embedding $\Pow S{\leadsto}\Pow\Dist S$ is as sets of point distributions, and for (2) the embedding $\Dist S{\leadsto}\Pow\Dist S$ is as singleton sets of distributions.

The full probabilistic/demonic model has been thoroughly explored in earlier work \cite{Morgan:96d,He:97,McIver:05a} and has an associated simple programming language \pGCL, for which it provides a denotational semantics.
\footnote{This approach is also similar to the work of Segala \cite{Segala:95b}, whose construction based on I/O automata appeared at about the same time as the workshop version of \cite{He:97}; and it has numerous connections with probabilistic/demonic process algebras as labelled transition systems that alternate between demonic- and probabilistic branching.}

Using \pGCL\ semantics, we can model our system as a while-loop of the form
\[
 \texttt{while s${\not\in}S_0$ do {\it ``choose \texttt{s'} according to $T(\texttt{s})$''}; s:= s' end}~,
\]
where \textit{``choose \texttt{s'} according to $T(\texttt{s})$''} is simply a \pGCL\ probabilistic/demonic assignment statement and the semantics of \texttt{while} is given as usual by a least fixed-point. 

An alternative, more recent approach is concerned with expected time to termination, and while-loops' semantics are given equivalently as limits of sequences of distributions \cite{Kaminski:16}. Either way, the resulting set of final distributions (non-singleton, if there is nondeterminism) comprises \emph{sub}-distributions, summing to no more than one (rather than to one exactly), where the ``one deficit'' is the probability of never escaping the loop. Proving \ACT\ then amounts to showing that all those sub-distributions are in fact full distributions, summing to one.

Our relying on well established semantics for demonic choice and probability together is the reason we do not have to construct a scheduler explicitly, as some approaches do: the scheduler's actions are ``built in'' to the set-of-distributions semantics.


\Section{Proof of the new rule for almost-certain termination}\label{s1351}

\Subsection{Proof of soundness}\label{s0839}
\begin{Theorem}{Soundness of \Sec{s1203}}{t1356}
The technique described in \Sec{s1203} is sound.
\Proof
Recall that the state space is $S$, that the termination subset is $S_0{\subseteq}S$ and that $S_\ast=S{-}S_0$ is the rest. The transition function $T$ is of type $S_\ast{\Fun}\Pow\Dist S$ and the variant $V$ is of type $S{\Fun}\NNReal$ with $V(S_0)=\{0\}$.

Fix some non-negative real number $H$ (for ``high''), and consider the subset $S_H$ of $S_\ast$ whose variants are no more than $H$, that is $\{s{\In}S_\ast\mid V(s){\leq}H\}$. By the non-increasing constraint on $p,d$ we have that for every $s$ in $S_H$ any transition decreases $V(s)$ by at least $d(v){\geq}d(H){=}d_H$ say, with probability at least $p(v){\geq} p(H){=}p_H$. Note that there does not have to be an actual $s$ in $S_\ast$ with $V(s){=}H$ for this condition to apply.

Now fix $s$ in $S_H$ with therefore $V(s){\leq}H$. The probability that $V$ will eventually become 0 via transitions from that $s$ is no less than $(p_H)^{\lceil H/d_H\rceil}$, since taking the probability-at-least-$p_H$ option to decrease $V$ by at least $d_H$, on every transition, suffices if that option is taken at least $\lceil\NF{H}{d_H}\rceil$ times in a row.

Since the above paragraph applies for all $s$ in $S_H$, the probability of transitions' escaping $S_H$ eventually is bounded away from zero by $(p_H)^{\lceil H/d_H\rceil}$ uniformly for \emph{all} of $S_H$. We can therefore appeal to the zero-one law \cite{Hart:83},\cite[Sec.\,6]{Morgan:96b},\cite[Sec.\,2.6]{McIver:05a}, which reads informally
\begin{quote}
Let process $P$ be defined over a (possibly infinite) state space $S$, and suppose that from every state in some subset $S_H$ of $S$ the probability of $P$'s eventual escape from $S_H$ is at least $\varepsilon$, for some fixed $\varepsilon{>}0$. Then $P$'s escape from $S_H$ is \AC: it occurs with probability one.
\end{quote}
Note that the zero-one law applies even if $S_H$ is infinite.

It is possible however that the escape occurs from $S_H$ not by setting $V$ to 0 but rather by setting $V$ to some value greater than $H$, i.e.\ occurs ``at the other end''. Because of possible nondeterminism, there might be many distributions describing the escape from $S_H$; but because we know escape is \AC, they will all be full distributions, i.e.\ summing to one. Let $\delta$ be any one of them.

Set $z{=}\delta_{S_0}$, i.e.\ so that the probability of indeed escaping to $V\kern-.2em{=}0$ is $z$. Then the probability of escaping to $V{>}H$ instead is the complementary $1{-}z$ for that $\delta$, and the expected value of $V$ over $\delta$ is at least $z{\times}0+(1{-}z){\times}H$, since the actual value of $V$ in the latter case is at least $H$. But by {\SMart}, we know that the expected value of $V$ when escape occurs from $S_H$, having started from $s$, cannot be more than $V(s)$ itself. So we have $(1{-}z)H\leq V(s)$, whence $z\geq1{-}\NF{V(s)}{H}$.

Now we simply note that the inequality $z\geq1{-}\NF{V(s)}{H}$ holds for any choice of $s,H$ and, in particular,
\marginpar{$\dagger$}
having fixed our $s$ we can make $H$ arbitrarily large. Thus $z$, the probability of escape to $V\kern-.2em{=}0$, i.e.\ to $S_0$, must be 1 for all $s$.
\footnote{\Label{n0925}A subtle issue here is that there might be $V{=}0$ states that $s$ can reach via all of $S_\ast$ but from which it is blocked because it must terminate when $V{>}H$ --- and our $z$ above does not take those into account. That is, the inequality wrt.\ $z$ might apply only to a \emph{subset} of the $V\kern-.2em{=}0$ states that $s$ can reach in the full system $S_\ast$. But the ``actual $z$'', i.e.\ for the full system, can only be greater still --- and so the result holds regardless.}
\end{Theorem}

\Subsection{Discussion of the rule and the necessity of its conditions}\label{s1905}
\Subsubsection{What happens if $V$ is not a super-martingale?}
Then \ACT\ could be be (unsoundly) proved e.g.\ for a biased random walker biased away from 0, say $\NF{1}{3}$ probability of stepping closer to zero and $\NF{2}{3}$ of stepping away. Setting its variant equal to its distance from zero satisfies {\Progress}, but not {\SMart}.

\Subsubsection{What happens without {\Progress}?}
Then a stationary walker would be compliant, satisfying \SMart\ but not \Progress. (Remember our \SMart\ does not require strict decrease: a stationary walker would satisfy it.)

\Subsubsection{Why not allow $V$ to go below 0?} In the proof we argued the expected value of $V$ on exit from $S_H$ would be at least $z{\times}0+h{\times}H$ --- but it could be much lower if an exit in the zero direction could set $V$ to a negative value.

In fact $V$ can be boundedly negative: we would just shift the whole argument up. But $V$ must be bounded below, otherwise the rule is unsound. Consider the ``captured spline'' example (in \Fig{f1216} of \Sec{s0937} below), and replace the 0-variants for escape by variants $-2(n+1)^2$. The $\nabla$ rule (defined in \Sec{s0807}) would now apply with $\nabla$ the constant function $-1$. For the current $p,d$ rule we could use the large negative escape-variants to increase the (positive) along-the-spline variants so that they became unbounded.

\Subsubsection{What happens when $V$ is bounded?}
Consider again the $\NF{2}{3}$--$\NF{1}{3}$ biased random walk. We can synthesise a (super-)martingale by setting $V(n){=}0$ when $n{=}0$ and solving for $\NF{V(n{-}1)}{3} + \NF{2V(n{+}1)}{3} = V(n)$ otherwise --- it gives the definition $V(n)=\NF{2^n-1}{2^{n-1}}$ with which \SMart\ is satisfied by construction. Then, since $V$ is injective, we can go on to define $p(v),d(v)$ to be the probability,decrease resp.\ actually realised by the process whenever its variant is $v$, appearing at first sight to satisfy \Progress\ trivially: set $p$ to be the constant function $\NF{1}{3}$ and $d(v)$ to be $2{-}v$ in this example. Both $p,d$ are non-increasing and strictly positive over variant values taken by the process.

But \Progress\ is not satisfied, because the functions $d,p$ must be defined and non-increasing over \emph{all} positive values $v$ and, in particular, not only over variant values actually taken by the process: that is, they must be defined even for values $v$ for which there is no $s$ with $v{=}V(s)$. In this example $d(v)$ decreases to 0 as $v$ approaches but never reaches 2, and so we cannot set a non-zero and non-increasing value for $d(2)$ itself. (In \Sec{s0937} a similar example is given where instead it is $p(2)$ that cannot be defined.)

The point in the proof at which this ``any $v$ whatsoever'' is used is marked by a marginal $(\dagger)$, where we let $H$ increase without bound. That $H$ does not have to be $V(s)$ for any ``actual'' $s$.

In summary: if $V$ is bounded but the values of ``actual'' $d$ (or $p$) are not bounded away from zero, then for any $H$ greater than all $V(s)$ there can be no non-zero value for $d(H)$ (or $p(H)$) and the proof fails.
\footnote{See \Thm{t1217} in \Sec{s0800} a for place where unbounded variant seems to be required.\par Defining $p,d$ everywhere, rather than only on ``actual'' $v$'s, is not a burden if the $v$'s are unbounded: define for example $\hat{p}(v')$ to be the infimum of $p(v)$ for all actual $v$'s with $v{\leq}v'$. Those extra values $\hat{p}(v')$ are never used, since there are no states with $V(s){=}v'$: just the existence of the extra values is enough.\par The only time this trick does not work is precisely the case we are discussing, where $v$ is bounded but $p(v)$ tends to zero.}

\Subsubsection{Why are $p,d$ functions of the variant rather than of the state?}
~\\Indeed they could have been defined as functions of the state (simply by composing them with $V$). In that case the non-increase conditions would become
\begin{quote}
If states $s,s'$ are such that $V(s){<}V(s')$ then $p(s'){\leq}p(s)$ and $d(s'){<}d(s)$.
\end{quote}
But we would have to add that $V$ over $S_\ast$ must either take only finitely many values or be unbounded, because we would then no longer be considering the $v$'s that correspond to no $s$. That conflicts with our ``purely local reasoning'' goal.

\Subsubsection{Why not simply require $V$ to be unbounded?}
For a finite state space $V$ cannot be unbounded; yet for finite state spaces a termination argument is (usually) easy. As our rule stands, termination for finite state-spaces is handled as part of the general argument, not as a special case.

%

%
\Subsubsection{Are there alternatives formulations of {\Progress}?}
~\\Yes: there are several alternatives.
\par The rule (\Sec{s1203}) uses {\Progress} in its proof (\Sec{s0839}) only to show bounded-away-from-zero escape from an arbitrary but bounded $V$-region $(0,H]$ that we called $S_H$. That is, starting from any $s$ in $S_H$ the probability of reaching eventually an $s'$ with either $V(s'){=}0$ or $V(s'){>}H$ is bounded away from 0, where the bound can depend on $H$. (It is {\SMart} that then converts that to \AC\ escape to $S_0$ alone, that is $V(s'){=}0$, by letting $H$ increase without bound.) Any other condition with the same force would do, and a significant programming-oriented example is given in \Sec{s0807}.

Another alternative, more suited to the situation where $S,T$ are laid out as a transition system or as a Markov process (but not so suitable for systems expressed as programs), is simply to require that the $V$-image of $S_\ast$ have no accumulation points. (An example of this kind of condition is found in \cite[Item (i)]{Kendall:1951aa}and \cite[Condition (2) \emph{proper divergence}]{Foster:1952aa}.) In that case the size of the set $V(S_H)$, i.e.\ the ``number of $V$'s'' in any region $(0,H]$, is required to be finite for any $H$. If the system is deterministic (or at least only boundedly nondeterministic)
\footnote{Both \cite{Kendall:1951aa,Foster:1952aa} deal only with deterministic systems, i.e.\ \emph{stationary} Markov processes.}
then if in every transition $V$ must decrease, by no matter how small an amount and with no matter how small a probability, escape from $(0,H]$ is assured because Zeno-effects cannot occur in a finite set: the $p(v),d(v)$ required in our rule (\Sec{s1203}) can be synthesised by taking minima over the whole (finite) set $(0,v]$, i.e.\ with $H{=}v$.

\Section{An equivalent rule based on parametrised strict super-martingales}\label{s0807}
\Cf{\Lem{l0835} and \Lem{l0926} need checking.}
Pursuing the theme of equivalent formulations of \Progress\ (mentioned just above), we give here an equivalent rule in which \Progress\ is removed altogether, and replaced by parametrically strict \SMart\ as follows:
\begin{equation}\label{e2005}
\parbox{0.9\textwidth}{
There must be a non-increasing strictly positive function $\nabla$ on the positive reals such that whenever we have $V(s){=}v$ for some $s,v$ and some $\delta$ in $T(s)$ then $\Exp_\delta V \leq v{-}\nabla(v)$.
\footnotemark}
\end{equation}%
\footnotetext{Note that $\nabla$ must be defined on \emph{all} the positive reals, not just on the variant values the process can actually take.}%
Call this formulation  the ``$\nabla$ rule'', and the original the ``$p,d$'' rule. Although the $\nabla$ rule is simpler to state than the $p,d$ rule, in practice the definition of $\nabla$ can be complicated, often the definitions of $p,d$ are more straightforward. The similarity of this rule with other strict super-martingale rules is clear: our condition is weaker (the  rule stronger) because we do not impose a uniform $\varepsilon$ across all of $S_\ast$.

\Ct{Have not checked the material from here to the end of \Sec{s0807}.}
We show first that the $p,d$ rule implies this $\nabla$ rule.
\begin{Lemma}{(Technical)}{l0835}
[Let $f$ be a non-negative function] over the non-negative reals, and let $y,y'$ be non-negative reals; let $\delta$ be a discrete distribution on the non-negative reals. Then
\begin{Equation}\label{e0955a}
 \Exp_\delta f \leq y \WideRm{implies} \delta_{\{x\mid f(x){<}y'\}}\geq1{-}\NF{y}{y'}~.
\end{Equation}%
That is, if $\delta$ guarantees that the expected value of $f$ is no more than some $y$, then for any $y'$ we have that $\delta$ is guaranteed with probability at least $1{-}\NF{y}{y'}$ to set $f$ to a value no more than $y'$.
\footnote{If $y'{\leq}y$ then of course this guarantee is vacuous.}
\Proof
Let $p$ be the aggregate probability that $\delta$ assigns to $\{x\mid f(x){\geq}y'\}$. Then, since $\delta$ is fixed, the smallest possible value of $\Exp_\delta f$  is $py'$, found by making $f$ itself as small as possible: that occurs when $f(x){=}0$ for all $x$ with $f(x){<}y'$ and $f(x){=}y'$ for all $x$ with $f(x){\geq}y'$. Thus $py'\leq\Exp_\delta f \leq y$, whence $p\leq\NF{y}{y'}$ and so the complementary $ \delta_{\{x\mid f(x){<}y'\}}$ is $1{-}p\geq1{-}\NF{y}{y'}$.
\end{Lemma}

\begin{Lemma}{Guaranteed decrease of variant}{l0926}
Let $V,S,T$ etc.\ be as above. Suppose for some state $s$ in $S_\ast$ we have that any $T$-transition is guaranteed to decrease the expected value of $V$ by at least some $\varepsilon{>}0$.

Then any $T$-transition is guaranteed with probability $p$ to decrease the \emph{actual} value of $V$ by at least $d$, where $d\Defs\NF{\varepsilon}{2}$ and $p\Defs\NF{d}{V(s){-}d}$.

\Proof
Let $\delta$ in $T(s)$ be a $T$-transition from $s$, and for \Lem{l0835} set $y=V(s){-}\varepsilon$ and $y' = V(s){-}\NF{\varepsilon}{2}$ and $f{=}V$. Then $\delta$ is guaranteed with probability at least
\[
 1{-}\frac{y}{y'} \Wide{=} 1-\frac{V(s){-}\varepsilon}{V(s){-}\NF{\varepsilon}{2}} \Wide{=} \frac{\NF{\varepsilon}{2}}{V(s){-}\NF{\varepsilon}{2}}
\]
to decrease $V$ by at least $V(s){-}y'=\NF{\varepsilon}{2}$.

So we let $d$ be $\NF{\varepsilon}{2}$ and $p$ be $\NF{d}{V(s){-}d}$.
\end{Lemma}

We can now conclude that the $\nabla$ rule implies the $p,d$ rule because if the $\varepsilon$ in \Lem{l0926} is a non-increasing but never-zero function of $V(s)$, then the $p,d$-values synthesised there are also non-increasing never-zero functions of $V(s)$. Non-increase of $d$ follows from the assumed non-increase of $\varepsilon$, and the non-increase of $p$ follows from increase of $V$ and non-increase of $d$.

\bigskip
For the opposite direction, that the $\nabla$ rule implies the $p,d$ rule, we again let $V,S,T$ etc.\ be as above. we will replace variant $V$ by $V' = f{\Comp}V$ where $f$ is a real-valued function that is
\begin{itemize}
\item non-decreasing,
\item strictly concave and
\item of non-increasing curvature.
\end{itemize}
That would be equivalently $f'{\geq}0$ and $f''{<}0$ and $f'''{\geq}0$, for which an example is logarithm.

Now for any state $s$ and $\delta$  in $T(s)$, we know that with probability at least $\hat{p}{=}p(V(s))$ the $\delta$-transition decreases $V(s)$ by at least $\hat{d}{=}d(V(s))$, and from \SMart\ we know that that $\Exp_\delta V\leq V(s)$. Then because of the concavity of $f$, the smallest value of $V'(s){-}\Exp_\delta V'$ occurs when $\delta$ sends exactly weight $\hat{p}$ to (possibly several) $s'$ with $V(s')=V(s){-}d$ and exactly weight $\hat{p}'$ to $s''$ with  $V(s'')=V(s){+}d'$ where $\hat{p}'$ is $1{-}\hat{p}$ and $\hat{p}\hat{d}{+}\hat{p}'\hat{d}' = 0$. (The construction of $\hat{p}',\hat{d}'$ is to make $\hat{d}'$ as big as possible while satisfying \SMart\ wrt. $V$.)

Because of $f$'s concavity, that smallest value of $V'(s){-}\Exp_\delta V'$ will be non-zero; and because the curvature is decreasing, and $p,d$ are non-increasing functions of $V(s)$, it will be non-increasing wrt.\ increasing values of $V(s)$; because $f$ in non-decreasing, that is equivalently non-increasing wrt.\ increasing values of $V'$.
\Cf{need some more rigour here\ldots}

\Section{Relation to the rule of Fioriti and Hermanns }\label{s0819}
Fioriti and Hermanns' rule \cite{Fioriti:2015} does not have our {\Progress} condition; instead they require uniform bounded-away-from-zero decrease of the expected value of the variant, that is with the same bound for the whole of $S_\ast$.

But in \Sec{s0807} we showed that our rule is equivalent to one without \Progress, i.e.\ where \SMart\ has been strengthened to the $\nabla$ rule at \Eqn{e2005} above. 

Fioriti and Hermanns' rule is then the special case of \Eqn{e2005} where $\nabla$ is the everywhere-$\varepsilon$ constant function. Furthermore, since that rule is complete for systems with finite expected time to termination, the result above means our proposed rule is also complete for that class.
\Cf{Mention here where we talk about ``Foster completeness''.}
But --as observed in \Sec{s1203}-- our rule also applies to the unbounded random walk, where the termination time is infinite.

For further discussions of completeness, see \Sec{s1031}.

\Section{Examples of termination and non-termination}\label{s1958}

\Subsection{Symmetric unbounded random walk (terminates)}\label{s0910}~\\
We mentioned in \Sec{s1203} that with variant the ``distance from 0'' and $p,d$ the constant functions $\NF{1}{2},1$ respectively the \ACT\ of the one-dimensional symmetric random walker is immediate. We also stressed our concern with source-level reasoning. Here we  illustrate such reasoning for a random-walk program:
\begin{Prog}
          s:= 1
          while s$\neq$0 do
              s:= s+1 $\PCF$ s-1
          end
\end{Prog}
Reasoning in Kozen's style \cite{Kozen:85} (here written in \pGCL\ \cite{Morgan:96d,McIver:05a}) would generate just these two elementary verification conditions for the proof-rule of \Sec{s1203}:
\footnote{In fact they are both equalities; but in general the inequalities shown are what must be verified. \Ct{Actually, have to account for $s{\neq}0$ somehow.}}
\[
 \begin{array}{cp{10em}r@{~~}c@{~~}l}
  \cdot & \multicolumn{4}{p{30em}}{The expected value of the variant does not increase:}\\
           & \SMart & s & \geq & \WP.(\texttt{s:= s+1 $\PCF$ s-1}).s \\[1ex]
  \cdot & \multicolumn{4}{p{30em}}{With probability at least $\NF{1}{2}$ the variant decreases by at least 1:}\\
           & \Progress & \NF{1}{2}[s{=}N] & \leq & \WP.(\texttt{s:= s+1 $\PCF$ s-1}).[s{\leq}N{-}1]
 \end{array}
\]
The \WP\ is the probabilistic generalisation of Dijkstra's weakest precondition \cite{Dijkstra:76,Kozen:85,Morgan:96d,McIver:05a}.

To allay suspicions that might be raised by the simplicity of the above, we ``uppack'' the reasoning used in the proof of \Thm{t1356}, showing in particular how the zero-one law contributes in this particular example. Without loss of generality we take the state-space to be the non-negative integers, start at position $s{=}1$ and show that eventually we will reach $s{=}0$.

Consider say the segment $1{\leq}s{\leq}100$ of the line, and the \emph{bounded} random walk within it, beginning (as we said above) at $s{=}1$. Since $s$ is decreased by $d{=}1$ with probability $p{=}\NF{1}{2}$ at every step, i.e.\ the \Progress\ property,  the walker's chance of moving to $s{=}0$ is at least $\NF{1}{2^{100}}$ for every $1{\leq}s{\leq}100$. Thus its escape from $[1,100]$ is \AC, whether that escape is high or low, and the expected value of $s$ when that happens will be $z{\times}0+(1{-}z){\times}100$, that is $100(1{-}z)$, where $z$ as before is the probability of escaping to $V{=}0$.

But the expected value of $s$ is constant at 1 (the \SMart\ property), no matter how many steps are taken, so that in fact $z{=}\NF{99}{100}$. That is, the probability that escape occurs to $s{=}0$ rather than to $s{=}101$ is $\NF{99}{100}$, establishing in any case that $s{=}0$ is reached from $s{=}1$ with at least that probability.

Now replay the argument within the segment $1{\leq}s{\leq}10^6$ instead. The walker's behaviour is not affected by the segment within which we reason --it does not ``know'' we are looking at $[1,10^6]$-- and it moves just as it did in the 100 case. But because we are thinking about $10^6$ this time, our conclusions are strengthened to ``\,escape from $s{=}1$ to $s{=}0$ with probability $1{-}\NF{1}{10^6}$\,''.

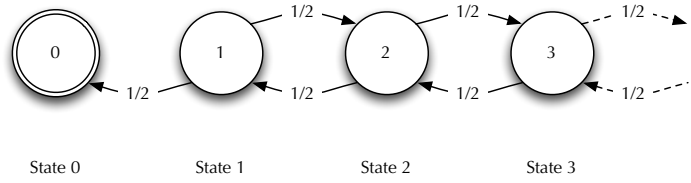
\begin{figure}
\ImageInTextBlock{15ex}{6ex}{0pt}{0pt}{12ex}{0.55}{Ex1312}

The $p,d$ version of our rule (\Sec{s1203}) establishes \ACT. The variant is the distance from 0 which, everywhere except 0 itself, is a (super) martingale that decreases by at least $d{=}1$ with probability at least $p{=}\NF{1}{2}$.
\caption{The unbounded symmetric random walk example}\label{f1312}
\end{figure}

\Subsection{Constant-bias random walk (non-terminating)}
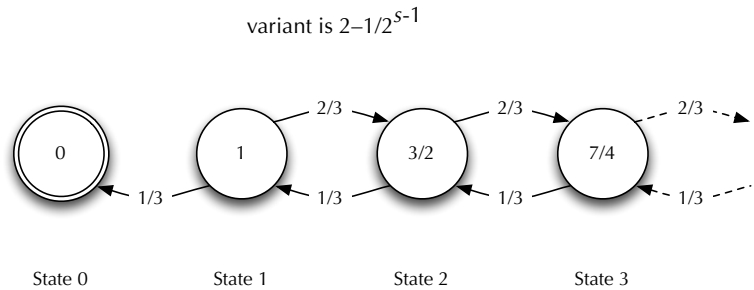
\label{s0904a}
\begin{figure}
\ImageInTextBlock{25ex}{6ex}{0pt}{0pt}{6ex}{0.6}{Ex1129a}

Here the walker has constant bias away from 0, and indeed termination is not \AC.

\medskip  Although \SMart\ is satisfied and we can define $p(v){=}\NF{1}{3}$, it is impossible to define a non-increasing function $d$ that gives a lower bound on the amount by which the variant decreases: the variant at State $s$ is $2{-}\NF{1}{2^{s-1}}$, bounded above by $2$ and forcing the non-increasing but strictly positive $d$ impossibly to satisfy $d(2){=}0$.

\caption{The constant-bias random walk example}\label{f1129a}
\end{figure}
In \Fig{f1129a} we have a one-dimensional random walk that does \emph{not} terminate \AC. If we synthesise a variant $V$ that is an exact martingale, as shown, we satisfy \SMart\ by construction. And its decrease occurs with probability (at least) $\NF{1}{3}$ everywhere. But because the variant is bounded, we cannot define a $d$ that satisfies \Progress, so our termination rule does not apply. (And \Sec{s0807} shows that the $\nabla$-rule does not apply either.) In \S\S\ref{s1207},\ref{s1730},\ref{ss0154} we see that in fact this walker does not terminate \AC.

\Subsection{Harmonic-bias random walk (terminates)}\label{s0904}
\begin{figure}
\ImageInTextBlock{25ex}{6ex}{0pt}{0pt}{6ex}{0.55}{Ex1129}

Here we use the $p,d$ version (\Sec{s1203}) of the proof rule: the expected value of the variant after a transition is equal to its actual value before (except at State 0, where our rule does not require it to be). But still this walker is strictly biased away from 0 at all positions, with that bias however decreasing towards zero with increasing distance. In spite of that bias, still its termination is \AC.

\caption{The harmonic-bias random walk example}\label{f1129}
\end{figure}
In \Fig{f1129} we see a biased one-dimensional random walk that still terminates \AC. The key point is that the bias decreases as distance from 0 increases, tending to ``symmetric'' in the limit.
\footnote{Although we constructed this example ourselves, we later found it in \cite[Sec.\,3(b)]{Foster:1952aa}.}

Here the variant is unbounded. (Compare \Sec{s0904a} just above, where the variant is bounded.) Condition \SMart\ is satisfied by construction; define $p(v){=}\NF{1}{3}$ everywhere; and define $d(v)$ to be $\NF{1}{s}$ where $s$ is the largest such that Harmonic Number $H_s$ is no more than $v$. This $d$ is non-increasing in $v$ and strictly positive.

An alternative proof of termination for this process is provided by the general techniques of \Sec{s0800}.

\Subsection{The ``tinsel'' process (terminates)}\label{s0918}
Here we exhibit  a process whose infinite stopping time is obvious from its construction. (The random-walk process (\Sec{s0910}) has the same infinitary property, but it is not so obvious.)

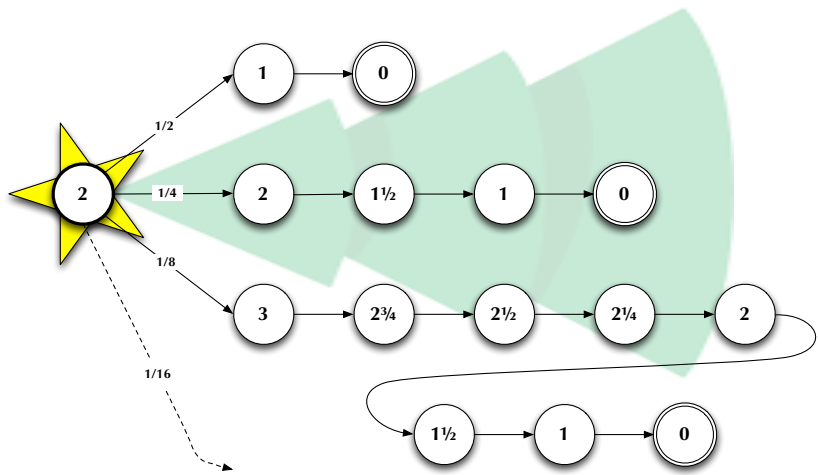
\begin{figure}
\ImageInTextBlock{45ex}{6ex}{0pt}{0pt}{2ex}{0.4}{Ex0829}

Variants in $(0,1]$ decrease by at least $1$ with probability at least $1$. In $(1,2]$ the (smaller) lower bound is $\NF{1}{2}$; in $(2,3]$ it's $\NF{1}{4}$; in $(3,4]$ it's $\NF{1}{8}$\ldots 

\medskip
The \SMart\ condition is satisfied trivially except at the root node, where the small calculation $1/2^1{+}2/2^2{+}3/2^3{+}\cdots = 2 \leq 2$ is needed to see that it is satisfied there too.

\medskip
The expected stopping time however is $\sum_{n\geq1} 2^n/2^n = \infty$. (We call it ``tinsel'' because it's like long ribbons hanging down from a tree.)

\caption{The ``tinsel'' process (rotated $90^\circ$)}\label{f0829}
\end{figure}

The root branches with probabilities $\NF{1}{2}, \NF{1}{4}\cdots,\NF{1}{2^n},\cdots$ to straight paths of length $2,4,\cdots,2^n,\cdots$ resp.\ each of whose contributions to the expected stopping time is therefore $(\NF{1}{2^n})/(\NF{1}{2^n})=1$. Since there are infinitely many children of the root, the expected stopping-time overall is infinite. See \Fig{f0829}, where the variant function for \ACT\ is shown.

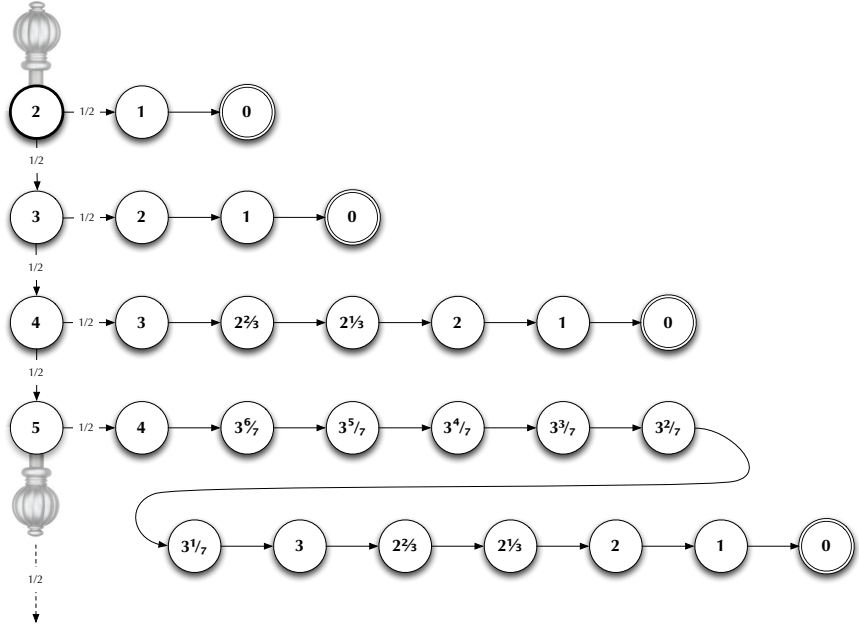
\begin{figure}
\ImageInTextBlock{60ex}{6ex}{0pt}{0pt}{2ex}{0.35}{Ex1344}

Variants in $(0,2]$ decrease by at least $1$ with probability at least $\NF{1}{2}$. In $(2,3]$ it's $\NF{1}{3}$; in $(3,4]$ it's $\NF{1}{7}$\ldots\ In $(s{-}1,s]$ it's at least $\NF{1}{(2^{s-2}{-}1)}$.

\medskip
The \SMart\ condition is satisfied trivially everywhere.

\medskip
The expected stopping time however is again $\sum_{n\geq1} 2^n/2^n = \infty$, as for \Fig{f0829}. (We call it ``curtain'' because many short runs hang down from a single long run.)

\caption{The ``curtain'' process (again rotated $90^\circ$)}\label{f1344}
\end{figure}

\Subsection{The ``curtain'' process (terminates)}
This variation on infinite stopping time begins with transitions that either move away from the root or ``drop down'' to ever longer straight runs. Again the stopping time is infinite but termination is still \AC. See \Fig{f1344}, where the variant function is shown.

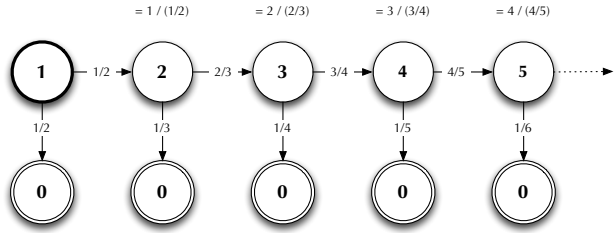
\begin{figure}
\ImageInTextBlock{25ex}{6ex}{0pt}{0pt}{12ex}{0.4}{Ex1215}

Each horizontal transition has probability of one minus the (vertcally downwards) escape immediately before it. Each variant $2,3,4,\cdots$ turns out to be the previous variant divided by its incident probability, establishing \SMart\ by construction. The successive probabilities of \emph{not} having escaped are corresponding prefixes of the infinite product $\NF{1}{2}{\times}\NF{2}{3}{\times}\NF{3}{4}{\times}\cdots$\, which tend to zero. Hence the variant increases without bound, proving that eventual escape is \AC.

\medskip
In general, if the product of the ``stay on spline'' probabilities tends to zero, the variants --the reciprocals of those prefix probabilities-- increase without bound.
\caption{The ``escaping spline'' process}\label{f1215}
\end{figure}

\Subsection{The escaping spline (terminates)}
Here we illustrate in \Fig{f1215} how our rule depends on the actual transition probabilities in an intuitive way, that a ``spline'' whose overall probability of being followed forever is zero gives a variant with which we can prove its termination. (Complementarily, if the probability of remaining in the spline is not zero then our rule does not apply, as we show in \Sec{s0937}.)

The states are numbered from $s{=}1$ at the left, and $V(s)$ is $s$ itself. The function $p(v)$ is $\NF{1}{v+1}$ and the function $d(v)$ is 1 everywhere: at state $s{\neq}0$ the variant is $s$, and with probability at least $\NF{1}{s+1}$ the value of the variant will decrease by at least 1. In fact, for most $s$ with $V(s){\neq}0$ the variant by much more than 1 --- the function $d$ gives only a lower bound for the actual decrease in the variant.

\Subsection{The captured spline (non-terminating)}\label{s0937}
In the example of \Fig{f1216}, based on  \cite[Sec.\ 2.9.1]{McIver:05a}, the process does not escape with probability one. If we applied the strategy of the escaping spline (\Fig{f1216}), we would choose variant $V(s)= \NF{2s}{s+1}$. It is a (super-)martingale because in general
\[
 \begin{array}{cl}
       & V(s{-}1){\times}0 + (1{-}\NF{1}{(s+1)^2}){\times}V(s{+}1) \\[2ex]
   =  & \NF{1}{(s+1)^2}{\times}0 + (1{-}\NF{1}{(s+1)^2}){\times}(\NF{2(s{+}1)}{s+2}) \\
   =  & (\NF{s^2{+}2s}{(s+1)^2}){\times}(\NF{2(s{+}1)}{s+2}) \\
   =  & (\NF{s^2{+}2s}{s+2}){\times}(\NF{2(s{+}1)}{(s+1)^2}) \\
   =  & \NF{2s}{s+1} \\[2ex]
   =  & V(s)~.
 \end{array}  
\]
The decrease function $d$ is trivial: we can set it to the constant 1, since the potential decrease is always \emph{at least} 1 with probability $\NF{1}{(s+1)^2}$.

But for $p(v)$ we  choose $\NF{(2{-}v)^2}{4}$, i.e.\ a value that is no more than $\NF{1}{(s{+}1)^2}$ when $v=\NF{2s}{s+1}$. Whatever that value is, it is clear that it approaches 0 as $v$ approaches 2, and so we will not be able to select a non-zero value for $p(2)$. As for \Sec{s0904a}, the results of \S\S\ref{s1207},\ref{s1730},\ref{ss0154} show that this process does not terminate \AC.
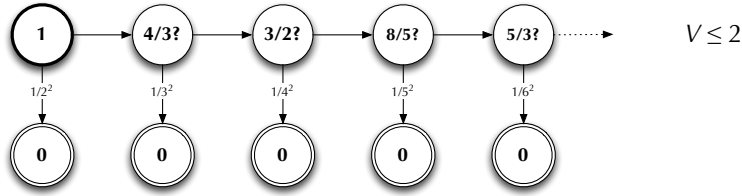
\begin{figure}
\ImageInTextBlock{20ex}{6ex}{0pt}{0pt}{12ex}{0.4}{Ex1216}

As before, each horizontal transition has probability (this time not shown) of one minus the escape immediately before it; and each (speculative) variant is the previous variant divided by its incident probability. The successive probabilities of \emph{not} having escaped are now corresponding prefixes of an infinite product $(1{-}\NF{1}{2^2})\times(1{-}\NF{1}{3^2})\times(1{-}\NF{1}{4^2}){\times}\cdots$\, which, unlike the earlier one of \Fig{f1215}, does not diverge: rather it converges to $\NF{1}{2}$. Hence eventual escape is with probability only $1{-}\NF{1}{2}=\NF{1}{2}$.

\medskip
Making  the variants the reciprocals of those cumulative escape probabilities, as in \Fig{f1215}, results in increasing variants bounded above by 2, which does not satisfy \Progress\ for $p(v)$ when for example $v{=}2$.

\medskip
In general, the strategy of Figs.\,\ref{f1215},\ref{f1216} works just when the successive ``not yet escaped'' probabilities tend to zero, since that is exactly when the variants, their reciprocals, increase without bound.
\caption{The ``captured spline'' process}\label{f1216}
\end{figure}

\Subsection{The two-dimensional random walk (terminating but not proved)}\label{s1732}

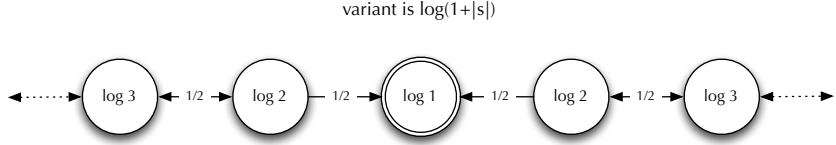
\begin{figure}
\ImageInTextBlock{12ex}{6ex}{0pt}{0pt}{2ex}{0.5}{Ex1312a}

Here we use the $\nabla$-version (\Sec{s0807}) of the proof rule: the expected value of the variant decreases by at least some fixed positive and non-increasing function of its current value: the expected decrease here is $\frac{1}{2}\log\frac{n^2}{n^2-1}$, a non-increasing function of $\log n$.

\caption{The unbounded symmetric random walk example}\label{f1312a}
\end{figure}

In \Fig{f1312a} we recall the one-dimensional random walk, but this time using a variant equal to the logarithm of (one plus) the walker's distance from the origin and a $\nabla$-style \Progress\ condition. (Compare \Fig{f1312} in \Sec{s0910}.) For better comparison with the two-dimensional version, we have made the walk unbounded in \emph{both} directions. It suggests that the two-dimensional walker could be treated with the variant being based on the logarithm of the walker's \emph{Euclidean} distance from the origin. Again using the $\nabla$ rule, we would have at least to show (something like) that for all integers $x,y$ we have
\[
 \begin{array}{c@{~~}l}
       &  \log ((x{+}1)^2{+}y^2) + \log ((x{{-}}1)^2{+}y^2) + \log (x^2{+}(y{+}1)^2) + \log (x^2{+}(y{{-}}1)^2) \\
   < & 4\log (x^2{+}y^2)~.
\end{array}
\]
Unfortunately, numerical calculations show that this inequality fails near the $|x|{=}|y|$ lines. It seems that the $\log$ function bends too much, is ``too concave''.

We therefore ``flatten things out a bit'' by trying a double-log $\log(\log(\cdot))$ instead, a function still concave but less so, and we have indeed shown by similar numerical calculations that the corresponding inequality
\begin{Equation}\label{e1736}
 \begin{array}{c@{~~}l}
       &  \Lgg (x{+}1)^2{+}y^2) + \Lgg ((x{{-}}1)^2{+}y^2) + \Lgg (x^2{+}(y{+}1)^2) + \Lgg (x^2{+}(y{{-}}1)^2) \\
   < & 4\Lgg (x^2{+}y^2)
\end{array}
\end{Equation}%
is satisfied for all integers $x,y$ with $|x|,|y|\leq 10,000$.
\footnote{We write $\Lgg$ for that function. Very close to the origin the it is undefined: but those cases can be adjusted manually.}

Our conjecture is that \Eqn{e1736} holds for all integers $x,y$ and, if it does,  it would establish termination for the two-dimensional symmetric random walk using a single variant function.
\footnote{As hoped, $\Lgg$ fails in the three-dimensional case.}

See \Sec{s1730} for evidence that there is a suitable variant function, even if it turns out not to be $\Lgg$.

\Section{Compositionality}
Following \cite{Fioriti:2015}, by ``compositionality'' we mean the synthesis of an \ACT-proof for a system that is composed of smaller systems for each of which we have an \ACT-proof already. For now, we study this only briefly.

Suppose we have a ``master'' system $M$ and a number of component systems $C_{1..N}$. System $M$ has at least $N$ termination states, at which its variant $V_M$ is therefore zero; and each component system has a designated start state $s_n$ where its variant function $V_n$ takes some value $v_n$. The composite system is then made by ``plugging in'' each component system's starting state $s_n$ to some termination state of $M$.

The systems in \Fig{f0829} (Tinsel) and \Fig{f1344} (Curtain) are examples of this, except that for them the number of component systems is infinite.

In Tinsel, the master $M$ is a single infinite branch leading with ever-decreasing probability ${1}/{2^n}$ to termination in exactly one step. Its component systems $C_{1..}$ are straight line processes each with stopping time $2^n{-}1$. The overall stopping time of the combination is infinite.

In Curtain the master $M$ is a straight-line system with a probability of $1/2$ of termination at each step; its expected stopping time is 2. The component systems $C_{1..}$ this time have termination times of $2^n{-}n$. Again the overall stopping time of the combination is infinite.

Although we did give termination proofs for these two systems, we cannot (i.e\ at the moment we do not know how to) \emph{synthesise} such proofs from the master's and the components' proofs when the number of components is infinite. But here is what we can do when the number of components is finite:

\begin{description}
\item[--]\makebox[0pt][r]{\$\hspace{2em}}Define $v_C$ to be the maximum over all $n$ of $v_n$, that is a number at least as great as the starting variant of any of the finitely many subsystems.
\item[--] Modify System $M$ by adding $v_C$ to its variant function $V_H$.
\item[--] Paste the starting node $s_n$ of each System $C_n$ into the appropriate termination state of $M$. (These are therefore no longer terminating states.)
\item[--] Set $\nabla_\sqcap$ for the new, single system to be the pointwise minimum over all $C_n$ and $M$ of their individual $\nabla_n$ and $\nabla_M$ functions.
\end{description}
If the individual systems satisfied the $\nabla$ rule with their separate $\nabla$ functions, then the composite system will satisfy the rule with the single $\nabla_\sqcap$ function. The use of finiteness was in two places:
\begin{description}
\item[--]\makebox[0pt][r]{\$\hspace{2em}}That the $v_C$ added to $V_H$ was finite. (It is a sup taken over all subsystems.)
\item[--] That the $\nabla_\sqcap$ is nowhere zero.
\end{description}

In Tinsel (\Fig{f0829}) and Curtain (\Fig{f1344}), having infinitely many components, the failure of synthesis occurs at the two points \textbf{\$}, because $v_C$ is infinite. In spite of that, as the examples show, we were able to find proofs ``by hand'' (i.e.\ not synthesised). Note however if a proof method were complete only for finite stopping-time systems, there can be no synthesis in these two cases: although all the component systems have finite stopping times, but the composite systems do not.

\Section{Related historical results on Markov chains}\label{s1031}

\Subsection{The work of Blackwell: random walks and radially symmetric trees}\label{s0800}

Blackwell \cite{Blackwell:55} gives a general technique for proving termination of a certain subclass of Markov processes, those moving both down \emph{and up} so-called ``radially symmetric trees''. (It also provides an independent proof of termination for our example \Sec{s0904}, the harmonic random walk.)

\begin{Definition}{Radially symmetric tree}{d2309a}
[A radially symmetric tree] is finitely branching, having the property that each node at depth $d$ has exactly $c_d$ children, where the root has depth zero and all the $c_0,c_1,\cdots$ are integers at least $1$.
\end{Definition}
A radially symmetric tree is infinite, and has no leaves.

\begin{Definition}{Random tree-walk}{d2309}
[A random walk] on a radially symmetric tree starts at any node and chooses uniformly to move either to its parent or to one of its children, thus with probability $\NF{1}{c_d{+}1}$ for a node at any positive depth $d$ along any of its connecting arcs. At the root, where there is no parent, the probability is instead $\NF{1}{c_0}$.
(Only the root has no parent.)

Termination occurs when the root is reached.
\end{Definition}

Radially symmetric trees are determined uniquely by their $c_d$'s, and examples include the following:
\begin{enumerate}[(i)]
\item  Each node has exactly one child, thus a single path starting at the root.
\item Each node has exactly two children, thus an infinite binary tree.
\item\label{i1348} Each node has either one or two children, according to the scheme that $c_d$ is two just when $d$ is a power of two (and thus is one otherwise).
\end{enumerate}

The first example generates the  one-sided symmetric random walk in one dimension, and it terminates with probability one (\Sec{s0910}). The second example  is (effectively) a collection of $\NF{2}{3}$--$\NF{1}{3}$ asymmetric random walks in one dimension, down the branches of the tree: its termination is not \AC\ (\Sec{s1905}).
The third, more complicated example is a binary tree with only infrequent splittings: we show in \Fig{f1839} below that it terminates, and quote a completeness result for such trees.

The Blackwell-style proof of \Itm{i1348}'s termination (a sanity check of our claim) follows \cite{Lessa:15}, and is ultimately based on Blackwell's \Thm{t1217} below. Note that it is an if-and-only-if:
\begin{Theorem}{Blackwell \cite{Blackwell:55}}{t1217}
[Let $p_n$ for $n{\geq}0$ be a sequence of probabilities,] with $p_0{=}1$, and consider the Markov matrix on the non-negative integers defined $M_{n,n+1}=p_n$ and $M_{n,n-1}=q_n=1{-}p_n$.

Then any Markov chain with this matrix will eventually reach $0$ almost surely \emph{if and only if} the equation $f(n) = q_nf(n{-}1) + p_nf(n{+}1)$ has no non-constant bounded solution.
\Af{This is quoted from Lessa's notes: I need to check the source.}
\footnote{Note however that our rule does not require $V$ to be unbounded. See \Sec{s1905}.}
\end{Theorem}

A corollary of \Thm{t1217} gives us a classification of \ACT\ for radially symmetric trees:
\begin{Corollary}{Lessa \cite{Lessa:15}}{c1223}
[Given a radially symmetric tree,] define variant function
\begin{Equation}\label{e1010}
 V(d) \Wide{\Defs} \sum_{0\leq i<d} \frac{1}{c_0{\times}\cdots {\times}c_i}
\end{Equation}%
where $d$ is a depth of some node in the tree and $c_d$ is the number of children for nodes at that depth. Every node at depth $d$ has variant $V(d)$, and this $V$ is indeed a super-martingale.
\Cf{ 
Check: Take any depth $d{>}0$, and calculate
\begin{Reason}
\WideStepR{}{All transitions from $d$ have probability $1/{1+c_d}$.}{
 1{\times}\frac{V(d{-}1)}{1+c_d}+c_d{\times}\frac{V(d{+}1)}{1+c_d}
}
\Space
\Step{$=$}{(\NF{1}{1+c_d}) ((\sum_{0\leq i<d-1} \frac{1}{c_0{\times}\cdots {\times}c_i})+c_d{\times}(\sum_{0\leq i\leq d} \frac{1}{c_0{\times}\cdots {\times}c_i}))}
\Space
\Step{$=$}{
    & (\NF{1}{1+c_d}) ((\sum_{0\leq i<d-1} \frac{1}{c_0{\times}\cdots {\times}c_i})+(\sum_{0\leq i<d-1} \frac{c_d}{c_0{\times}\cdots {\times}c_i})) \\[1ex]
 + & (\NF{1}{1+c_d}) (\frac{c_d}{c_0{\times}\cdots{\times}c_{d-1}}+\frac{c_d}{c_0{\times}\cdots{\times}c_d})
}
\Space
\Step{$=$}{
    (\sum_{0\leq i<d-1} \frac{1}{c_0{\times}\cdots {\times}c_i})
 + (\NF{1}{1+c_d}) (\frac{c_d}{c_0{\times}\cdots{\times}c_{d-1}}+\frac{1}{c_0{\times}\cdots{\times}c_{d-1}})
}
\Step{$=$}{
    (\sum_{0\leq i<d-1} \frac{1}{c_0{\times}\cdots {\times}c_i}) + \frac{1}{c_0{\times}\cdots{\times}c_{d-1}}
}
\Step{$=$}{
    (\sum_{0\leq i<d} \frac{1}{c_0{\times}\cdots {\times}c_i})
}
\Step{$=$}{V(d)~.}
\end{Reason}
} 

The random walk on this tree, defined as in \Def{d2309}, terminates everywhere if and only if $V(d)$ is unbounded as $d{\rightarrow}\infty$.
\Proof
\Cf{I'm not sure what's being proved here: is it Lessa's result in general, or just our \Fig{f1839}?}
\end{Corollary}

Lessa's proof of \Cor{c1223} uses Blackwell's theorem \Thm{t1217}; but our variant rule here provides an independent proof for \Fig{f1839}, i.e.\ without Blackwell's theorem. More significantly however,  Blackwell's theorem provides us with a completeness result for using our variant rule, at least for one-dimensional random walks.
\Cf{ 
Spell the completeness comment out more precisely. Something like this?
\begin{quote}
 Make the variant according to \Eqn{e1010} in the special case where $c_i$ is always 1, which satisfies \Thm{t1217}'s criterion by construction. Since that variant is unbounded, there are no variant-accumulation points and so the functions $p(v)$ and $e(v)$ can be defined by taking appropriate minima over all $v'{\leq}v$.
\end{quote}
} 
%

\begin{figure}
\ImageInTextBlock{60ex}{6ex}{0pt}{0pt}{-0.3em}{0.3}{Ex1839}

Every note at depth $2^n$ has two children; all others have one child. Transitions from a node are uniformly distributed over its arcs, thus $\NF{1}{3}$ for each of two children and $\NF{1}{3}$ up, and otherwise $\NF{1}{2}$ up and $\NF{1}{2}$ down.

\medskip
The variant function generated according to the scheme of \Eqn{e1010} is
\[\newcommand\Z[2]{\begin{array}[t]{ll}&#1\\+&#2\end{array}}
 \begin{array}{|p{17ex}|r|r|r|r|r|r|r|r|r|}
 \hline
 ~depth $d$& 0&1&2&3&4&5&6&7&8 \\\hline
 ~$c_d$& 1&2&2&1&2&1&1&1&2 \\\hline
 ~$V(d)$ from \Eqn{e1010}&0&\NF{1}{1}&\Z{1}{\NF{1}{1{\times}2}}&\Z{\NF{3}{2}}{\NF{1}{2{\times}2}}
  &\Z{\NF{7}{4}}{\NF{1}{4{\times}1}}&\Z{2}{\NF{1}{4{\times}2}}&\Z{\NF{17}{8}}{\NF{1}{8{\times}1}}
  &\Z{\NF{9}{4}}{\NF{1}{8{\times}1}}&\Z{\NF{19}{8}}{\NF{1}{8{\times}1}}\\\hline
 ~simplified &0&1&\NF{3}{2}&\NF{7}{4}&2&\NF{17}{8}&\NF{9}{4}&\NF{19}{8}&\NF{5}{2}\\\hline
 \end{array}
\]
At Depth 4, for example, we have $\NF{1}{3}{\times}\NF{7}{4}+\NF{2}{3}{\times}\NF{17}{8} = 2$, thus satisfying \SMart\ at that position. Because the variant at depth $2^d$ is $1+\NF{d}{2}$, i.e.\ increases without bound, it is straightforward to construct functions $p,d$, showing that this process terminates.
\caption{Blackwell's radially symmetric tree}\label{f1839}
\end{figure}

\Subsection{Blackwell's completeness result}\label{s1207}
Blackwell's work \cite{Blackwell:55} on classifying recurrence in Markov processes suggests how we might understand the coverage of our new rule. He considers Markov processes with countable state spaces and stationary (i.e.\ fixed) transition probabilities, and shows that such processes have essentially unique structures of recurrent and transient sets. We now give a summary, using (partly) Blackwell's terminology as well as what we have used elsewhere in the paper.

Let $C$ be a subset of the state space, and fix some initial state $\hat{s}$. Say that $C$ is \emph{almost closed} (wrt.\ that $\hat{s}$) iff the following conditions hold:
\begin{enumerate}
\item The probability that $C$ is entered infinitely often, as the process takes transitions starting from  $\hat{s}$, is strictly greater than zero and
\item If $C$ is visited indeed visited infinitely often, starting from $\hat{s}$, then eventually it remains within $C$ permanently.
\end{enumerate}

Say further that a set $C$ is \emph{atomic} iff $C$ does not contain two disjoint almost-closed subsets.

Finally, call a Markov process \emph{simple atomic} if it has a single almost-closed atomic set such that once started from $\hat{s}$ it eventually with probability one is trapped in that set. We then have

\begin{Theorem}{Corollary of Blackwell's Thm.\,2 on p656) \cite{Blackwell:55}}{t1125}~\\
A Markov process is simple atomic (as above) just when the only bounded solution of the equation $\Exp_{\delta}V=V(s)$, that is Blackwell's Equation (his 6), stating (in our notation) that $V$ is exact, neither super- nor sub, is constant for all $s$ in $S_\ast$ and $\delta$ in $T(s)$.
\end{Theorem}

We adapt the above to our current situation as follows. As above, we fix a starting state $\hat{s}$, and we collapse our termination set $S_0$ to a single state $s_0$, adjusting $T$ accordingly and in addition making $T$ take $s_0$ to itself. We then assume that the probability of $\hat{s}$'s reaching $s_0$ is one. We now note:
\begin{enumerate}[(1)]
\item Our termination set  $\{s_0\}$ is almost-closed and atomic, because
\begin{enumerate}[(i)]
\item almost closed: Our process reaches $s_0$ with non-zero probability (in fact we assumed with probability one) and, once at $s_0$, it remains there.
\item atomic: Our set $\{s_0\}$ has no non-empty subsets.
\end{enumerate}
\item We now recall that in fact $s_0$ is reached with probability one, so that the whole process is simple atomic.
\item From our \Thm{t1125} we conclude that the only possible non-trivial variant is unbounded.
\end{enumerate}
Thus --in summary-- we have specialised Blackwell's result to show that if we have a non-trivial exact variant that is bounded, then the process \emph{does not} terminate \AC. This is a  result in the style of Chatterjee et al.\ \cite{Chatterjee:16}. (See \Sec{ss1958}.)
%
%
%
%
%

\Subsection{The work of Foster: completeness}\label{s1730}
Foster \cite{Foster:1952aa} gives a characterisation of Markov processes for which a technique like ours is guaranteed to work. A significant example is the two-dimensional symmetric random walk, supporting our conjecture in \Sec{s1732}.

Because Foster's paper seems quite technical, we give here a ``translation'' into our own terms. His equations will be referred to as (F1) etc.\ and his sections as \S F1 etc.

\Cf{We need to check whether Foster's reference Kendall(3) has proved our rule already. No problem if so (since no-one else seems to know that); but we should acknowledge it if so. Our angle would then be ``to have embedded it in a programming language, and to have given a programming-logic proof'' (when eventually we do that). And of course we do nondeterminism, while he does not.
\par The reason I think he might have done it is that he seems to prove that if you have a (non strict) super-martingale without accumulation points, on a countable state space, then with probability one the process must eventually be ``captured'' in some finite subset (his $C$). That being so, if the \emph{only} capturing subset is $S_0$, as it often is, he would have proved our result. On the other hand, does he have weird conditions like Foster's (F7,8)? \ATD\ Do you agree? Can we get Kendall(3)?}

\Subsubsection{--- \S F1}\label{s0751}
We assume that our state space $S$ is countable, enumerated $s_{0,1,\cdots}$ with the termination subset $S_0$ being just a single point $\{s_0\}$, and we extend our transition function $T$ to all of $S$, i.e.\ not just over $S_\ast$, by making it take $s_0$ to itself. The enumeration should correspond roughly to ``being further from $s_0$'', which is made precise in conditions (F6) and (F8).

Foster is concerned with conditions for the existence of a super-martingale (F1) variant function $V$ from $S$ to the non-negative reals (F3), where $V$ is unbounded without accumulation points (F2).

We assume that $T$ is deterministic, and thus specialise it to be of type $S$ to $\Dist S$ rather than to $\Pow\Dist S$.

\Subsubsection{--- \S F2}
The ``limit'' of taking transitions forever is defined to be $T^*$, say, using the ``Cesaro average''
\footnote{\texttt{http://www.sciencedirect.com/science/article/pii/0304414977900321~.} \par\Ct{I am not \emph{absolutely} sure that Cesaro is what his $(C,1)$ refers to, though it seems very likely, since I have not yet found the exact notation ``$(C,1)$'' in any other literature. Foster's reference to Feller doesn't match the edition we so far have access to.}}
that avoids the problem of recurrence when considering simply $T^0$, $T^1$ etc.\ composed in the Markov style.

But (F4) is not as simple as it looks.
\Cf{Until we find a proper reference to $(C,1)$ we cannot be absolutely sure.}
It seems to imply that there is no infinite chain of transient states (such as in the ``spline'' examples). For if there were, the mass travelling down the chain would be ``lost'' in the Cesaro average, and the sum would not be one. This turns out to be important in the discussion of ($4'$) below.

Kendall's \cite{Kendall:1951aa} earlier result is
\begin{quote}
If there is a variant as in \Sec{s0751}, then $T^*$ takes any starting state to a full distribution on $S$ (i.e.\ not partial), and there is a \emph{finite} subset $C$ of $S$ from which $T$ does not escape.
\end{quote}
Foster then explains that the current paper's purpose is to explore the opposite implication to Kendall's \cite{Kendall:1951aa}, i.e.\ that, under ``certain weak additional assumptions'' on $T$, if there is such a (finite) subset $C$  as above then there is a $V$ satisfying the conditions of \Sec{s0751}.

His additional assumptions include (by implication) that $C$ is reached with probability one from anywhere in $S$, his (F$4'$), because he argues that (F4,F5,F6) together imply (F$4'$). That looks at first like the zero-one law. But note that (F6) does not bound the probability of escape away from zero: it merely says that it is not zero, and that is not usually enough. Together with (F4) though it suffices,
\Cf{I think}
because (F4) seems to say any transient state (even if there are infinitely many) must be visited infinitely often (since otherwise the mass moving among the transient states would be ``lost'', as suggested above).

His additional assumptions are then
\begin{description}
\item[(F6)] From any state in $S_\ast$ there is a non-zero probability of reaching $S_0{=}\{s_0\}$ eventually.
\item[(F7)] From any state in $s_i$ in $S_\ast$ there is for any $j{>}i$ a non-zero probability of reaching any $s_j$ eventually. Note that $s_j$ is in $S_\ast$ also.
\item[(F8)] There is a single probability $\delta{<}1$ for the whole system such that for any $N$ there is an $i$ such that for all $j{\geq}i$ the state $s_j$ cannot reach $C$ within $N$ steps and with probability at least $\delta$. As he says, it's a ``remoteness'' condition, intuitively mandating that the higher the $i$ the longer it takes $s_i$ to reach $C$ with some fixed-beforehand probability $\delta$.
\par He notes that because of (F$4'$) the $N$ (which depends on $i$) is finite: from $s_i$ you must get to $C$ eventually with probability at least $\delta$ because, in fact, you get there with probability 1.
\end{description}

\Subsubsection{--- Statement of Theorem F2, and its application to the two-dimensional symmetric random walk}
~\\

Recall that $S$ is assumed to be countably infinite. Foster's Theorem F2 reads
\begin{quote}\it
If $T$ satisfies conditions (F4--8), then there is a variant function $V$ on $S$ that satisfies (F1,F3), i.e.\ that it is a non-negative super-martingale and (F2)  that it tends to infinity as the state-index tends to infinity.
\end{quote}
We note that condition (F2) implies that $V$ is without accumulation points.
\Cf{Check that this really is equivalent.}

The implications of this theorem seem to be e.g.\ that there must be a variant in our style for the two-dimensional symmetric random walk, even if it has not (yet, as far as we know) been given in closed form. We check the conditions one-by-one:
\begin{description}
\item[(F4)] This is (apparently) replaced by (F$4'$).
\item[(F\boldmath$4'$)] The probability of reaching $S_0$, that is the origin, is one everywhere.
\item[(F5)] Once you are at the origin, you do not leave.
\item[(F6)] Every state in $S_\ast$ can reach $S_0$ with non-zero probability.
\item[(F7)] Here we need an enumeration of the states: Foster uses the Manhattan distance, which makes nested ``diamonds'' . But since \emph{every} state in $S_\ast$ can reach every other state, in fact we do not need the enumeration yet.
\footnote{Foster's (weaker) condition requires only that each state can reach every \emph{higher-enumerated} state. \Ct{And, as I remark later, I don't see what that isn't simply \emph{some} higher-enumerated state.}}
\item[(F8)] Any state at Manhattan distance $N$ cannot reach the origin at all until Step $N$, and so $\delta{=}0$ should do for this provided we assign higher indices to higher-Manhattan-distance states, which is what Foster does in \S F3.
\end{description}

Applying Theorem F2 then gives us a super-martingale $V$ such that $V(s_i)$ tends to infinity as $i$ tends to infinity, which means that $V(s)$ tends to infinity as $s$ gets Manhattan-further from the origin, given the indexing that we (i.e.\ Foster) have chosen.

To show that our rule applies, we need however to establish a progress condition. (See our earlier remarks about alternative progress conditions, in \Sec{s1905}.) First define $p(v)$ to be $\NF{1}{4}$ for all $v$. Then for $d$, first consider the subset $S_{\leq v}$ of $S$ comprising all those $s$ with $V(s){\leq}v$. Because of (F2) the $V$-image $V(S_{\leq v})$ of $S_{{\leq}v}$ must be finite; so set $d(v)$ to be the minimum non-zero distance between any two of them, that is $(\Min\, V(s'){-}V(s) \mid s,s'{\in}V(S_{\leq v})\land V(s'){>}V(s))$.
\Cf{This part of the argument does not depend on the 2dRW; we should move it into the ``Are there other progress conditions?''\ paragraph.}

Thus there is guaranteed to be a $V$ satisfying \SMart\ and \Progress\ that establishes termination for the two-dimensional symmetric random walk (\Sec{s1732}) --- even if we don't know what it is in closed form.  Foster's general proof is by construction, and we sketch it in \App{a1038}.

\Subsubsection{--- Why Theorem F2 does not synthesise a variant for the \\ \emph{three}-dimensional symmetric random walk}
~\\

Foster remarks \cite[p. 590]{Foster:1952aa} that synthesis cannot succeed for the three-dimensional random walk (since it is known that it is not \ACT); but he does not say which of his Theorem F2's conditions are not satisfied.

Clearly his (F4') is not satisfied (that the process is \ACT); but that is a derived condition, a consequence of his original (F4--8), and so it is fair to ask which of those original conditions fails in three dimensions. Furthermore, his synthesis procedure is well defined whether the process satisfies \ACT\ or not, and so we can therefore ask as well what is wrong with the $V$ it synthesises for the three-dimensional random walk, in our terms.

\smallskip
For the first, it is condition (F4) that must fail. The process satisfies (F5), that the process is trapped at the origin, and (F6), that the origin is accessible from every point, and (F7), that every point is accessible from every other (except the origin). And it seems likely that it satisfies (F8), since by numbering the states in rings it can be arranged that higher-numbered states have arbitrarily high first-arrival times at the origin.

Failure of (F4) means that there is some bounded away from zero probabilistic mass that follows an infinite (not looping) path through the state space: that is the only way in which the Cesaro limit can ``lose mass'', making (in Foster's notation) the sum $\sum_{j=0}^\infty\pi_{ij}$ strictly less than one.

\smallskip
For the second, the problem with the synthesised $V$ is that it is bounded, a failure of condition (F2). Item \ref{i0727}.\ in \Sec{ss0154} below shows that in that case our condition \Progress\ cannot be satisfied for that $V$.


\Subsection{A modern alternative to Blackwell's completeness argument}\label{ss0154}
The result of \Sec{s1207} can be obtained much more directly using the program semantics of this paper, and an argument in the style of \Thm{t1356}.

In \cite[Lem.\,7.3.1]{McIver:05a} we show (using the terminology here) that if $V$ is a sub-martingale and is bounded,
\footnote{Note we say \emph{sub}-martingale, i.e.\ that the expected value of $V$ can increase.}
\Af{Do you mean Martingale: the result is not true if we allow a strict expected decrease.}
\footnote{In that part of \cite{McIver:05a} we are working an invariant, here our $V$, that is bounded by 1. That loses no generality, since any bounded $V$ can be divided by its least upper bound without disturbing its sub-martingale property.}
then if escape from $S_\ast$ is \AC\ (i.e.\ the loop terminates with probability one), the expected value of $V$ on $S_0$ (i.e.\ on the states where the loop-guard is false) is at least the actual value of $V$ in the initial state (of the loop).
\footnote{See \Sec{s1728} for a pr{\'e}cis of this loop rule.}

Now if the initial state is in $S_\ast$ then the value of $V$ there is strictly positive; yet if escape occurs with probability one, the expected value of $V$ on termination will be 0, since it is confined to $S_0$. That contradiction means that we cannot have sub-martingale $V$ be bounded and still terminate almost surely.

\bigskip
Here are some remarks on the relation between our argument and Blackwell's \Thm{t1125}. Blackwell states that a process is \ACT\ just when its only bounded exact martingale is constant; our result just above states that an \ACT\ process cannot have a bounded sub-martingale.
\Af{For you, Annabelle, as a double check. \Cx I am not reliable here: I too much want it to be right, and so am likely to miss something.}
\begin{enumerate}
\item \textit{What role does the ``or is constant'' criterion, from Blackwell's theorem, play in our argument?}\\
Because we require $V$ to be zero on $S_0$, a constant $V$ for us would be zero on all of $S$, meaning that $S_\ast$ was empty (since $V$ must be strictly positive there). So we should add to our result ``unless $S_\ast$ is empty.''
\item[]
\item \textit{Where do we use in our argument that $V$ is bounded? \\ How does our argument fail if we don't?}\\
We use it in our appeal to \cite[Lem.\,7.3.1]{McIver:05a}, where in fact $V$ is assumed to be between 0 and 1. If $V$ is simply bounded above (but not by one, necessarily), then it can easily be brought within range by scaling. If $V$ is unbounded, however, it cannot be brought within range that way.
\par An easy counter-example is the symmetric random walk starting at 1 and aiming to reach 0. The variant ``distance from 0'' is an exact martingale, and has value 1 on initialisation. But it is unbounded, and so the conclusion ``its (expected) value on termination is at least its starting value'' is false. Indeed on termination its (actual) value has decreased from 1 to 0.
\item[]
\item \textit{What's an intuitive (and easy to understand now, in retrospect) reason that our conclusion is ``obvious'', without appealing either to Blackwell or to \cite{McIver:05a}?}\\
Think of the variant value $V$, initially concentrated at $s$, as being gradually ``dissipated'' whenever some probabilistic weight escapes to $S_0$. Since $V$ is zero there, the sub-martingale property requires that $V$ increase, to compensate, on the remaining probabilistic weight still within $S_\ast$. But because $V$ is bounded, that increase cannot go far enough --- it eventually must stop. And that means that some probabilistic weight remains trapped within $S_\ast$.
\item[]
\item\label{i0727}  \textit{Our rule \Sec{s1203} proves \ACT\ given \SMart\ and \Progress\ for some $V$. Yet above we show that if $V$ is a sub-martingale and bounded, then \ACT\ cannot hold. Does that mean that \Progress\ cannot hold for any bounded, non-zero \emph{exact} martingale?}\\
Yes, it does mean that. If $V$ is bounded, then when the process is at (or sufficiently near) $V$'s upper bound either
\begin{enumerate}
\item The function $d$ must be arbitrarily small (tending to 0), and so it cannot be non-increasing with respect to a non-zero $d$-value  taken above $V$'s upper bound (for which see \Fig{f1129a}), or
\item The function $d$ is bounded away from zero, in which case the expected value of $V$ must strictly decrease, thus not realising an \emph{exact} martingale.
\end{enumerate}
\end{enumerate}

\Ae{
I can't remember who calls things sub/super Martingales, but this is what I understand from what you have written:

To be proved: If $\Exp_\delta(V) = V(s)$ for all $s$ in $S_\ast$ and escape from $S_\ast$ is \AC, then in fact $V(s) = V(s_0)$.

Here we must have:

\begin{enumerate}[(a)]
\item $S_0= \{s_0\}$, or $V(s_0) = V(s_0')$ for all $s_0, s_0' \in S_0$, because otherwise we could have $s:= s_0~ { }_{1/2}\!\!\oplus s_0'$, but $s_0, s_0'$ assigned different values by $V$.  \C{But, for us, $V$ must assign 0 to all elements of $S_0$, however many there are. So the previous sentence is not a worry.}
\item We are only talking about \emph{equality} i.e. martingales, so not super and not sub, because we could just have $V(s) = 1$ for all $s\in S_\ast$, and $V(s) = 0$ for all $s \in S_0$. In a system where everything reaches $S_0$ eventually this is a sub (super??)-martingale, i.e. $\Exp_\delta(V) \leq V(s)$ --- it is bounded but not constant. There is no contradiction either: the (eventual) expected value at $s$ is $0$ everywhere and that is fine.
\item So I think that if your conclusion reads: ``we cannot have an exact Martingale that is bounded and not-constant in a system that terminates almost surely'' then it works. But I think that's what you mean...
\end{enumerate}

To answer your questions: 
\begin{enumerate}
\item Where do we use the ``or is constant'' from Blackwell?\\\\
I think somehow this has got lost in the super/sub business. Your proof shows that the eventual expected value is $V(s_0)$, which
means that it can't be an equality Martingale and be not-constant.\\
\C{I think Blackwell imposes his "is a constant" constraint on the whole system, does he not? Thus in our case "is a constant" would mean that $V$ was 0 everywhere, on $S_0$ and $S_\ast$ both. But we insist that $V$ is positive on $S_\ast$, so ``is a constant'' for us implies that either $S_0$ or $S_\ast$ is empty.}

\item Where do we assume that $V$ is bounded? How does the proof fail if we don't?\\\\
 I think that boundedness applies because you are using some convergence result which might be something
like  $\lim_{n\rightarrow \infty}\Exp_{T^n(s)}V = V(s)$ if we have an exact Martingale. In the 1-dimensional random walk that terminates at $0$ presumably that
doesn't hold, because otherwise we'd deduce that $n=0$ for all $n$, right? But presumably convergence does hold if $V$ is bounded.\\

\item What's an intuitive (and easy to understand now, in retrospect) reason that this theorem is ``obvious'', without appealing either to Blackwell or to \cite{McIver:05a}?\\\\

I think the intuitive answer is that (provided the convergence works out) if you have the exact Martingale, then the only
way the average is invariant is if the $V$ takes the value of $V(s_0)$ (and all the $s_0$'s have the same value!).

\C{I was thinking along these lines (and requiring only sub-martingale, i.e.\ that the expected value of $V$ can increase). If the whole thing ends up in $S_0$, then the expected value of $V$ has ``disappeared'' --- it was $V(\hat{s})$, where $\hat{s}$ is the initial state; but now it's zero. Where has it gone? Well the only place it can be is as the $V$ of some state $s'$ still in $S_\ast$ but with zero probability, and where that $V$ is ``infinite'' --- with $0{\times}\infty$ conveniently equalling exactly the missing weight. I know that's hand-wavy: but that's all I was asking for. But the other reason is that our $I{\leq}T$ loop rule does not work if the invariant (there $I$, here $V$) is unbounded. We have counter-examples for that\ldots}
\end{enumerate}
}

\Subsection{General comparison with refutation methods}\label{ss0829}
Blackwell's result \Thm{t1125} says that a Markov process is atomic and simple if and only if all exact martingales are constant or unbounded.
We showed (in our terms) that when a program terminates with probability 1, the termination set implies the program is atomic and simple (as a Markov Process). Then, using Blackwell's,  result we are able to conclude that all exact martingales are constant or unbounded. In an independent proof (the one-liner \Sec{ss0154}) we can show this directly without going through Blackwell, namely that if a program has a non-constant exact martingale then it can't terminate with probability 1.

Chatterjee et al. also look at repelling super-martingales to refute almost termination. Their \emph{Theorem 6} uses an $\varepsilon$-repulsing super-martingale with $\varepsilon{>}0$ to refute almost sure termination. Their \emph{Theorem 7} uses an $\varepsilon$ repulsing super-martingale with $\varepsilon\geq0$ to refute finite expected time to termination: i.e. to refute finite expected time to termination only a martingale is required.  

Our result in \Sec{ss0154} implies a new refutation certificate for programs: if the martingale is finite and non-constant it actually refutes termination with probability 1,  not just finite expected time to termination.

For example, if we consider the one-dimensional random walker \Sec{s0910} it has an exact \emph{unbounded} martingale, and therefore our rule shows that it terminates with probability $1$. The walker in \Sec{s0904a} has an exact \emph{bounded} martingale, and this we can conclude does not terminate with probability $1$. In both cases  Chatterjee's Theorem 7 would deduce that neither have finite expected time to terminate.

\Af{Double check that we can turn a martingale into a repulsing super martingale.}

\Section{Conclusion}
Our overall aim is (has always been) to allow and promote rigorous reasoning at the level of program source-code. In this paper we have proposed a new rule, combining earlier ideas of our own with important innovations of others, and have attempted to formulate it in a way that indeed is will turn out to be suitable for source code.

That is, we hope that as an extension of what's here we will be able to formulate these rules in the program logic \pGCL, or similar; and if the techniques are further extended subsequently, we would hope to do the same for those too.

Program logic also provides a rigorous setting not only for use of the rules but also for their proof in the first place. Although we did not use program logic here, for our proofs, we believe it would be possible e.g.\ in the style of \cite{McIver:05a}.

Finally, we have left an intriguing open question: is there an elementary variant for the two-dimensional random walk? Foster \cite{Foster:1952aa} shows that there is such a variant, but he does not give it in closed form. We conjecture that $\Lgg$ suffices, but have only verified that experimentally. Will we ever be able to set as a student exercise
\begin{quote}
Assuming the properties [\,\ldots\,] of the function [\ldots], use probabilistic assertions in the source code of the following program to prove that it terminates with probability one for any initial integers \texttt{X,Y}:
\begin{Prog}
          x,y:= X,Y
          while x$\neq$0 $\land$ y$\neq$0 do
              x,y:=     x+1,y
                     $\PC{}\,$ x-1,y
                     $\PC{}\,$ x,y+1
                     $\PC{}\,$ x,y-1
          end$~,$
\end{Prog}
\end{quote}
where iterated $\PC{}$ is shorthand for uniform choice (in this case $\NF{1}{4}$ each).

\section*{Acknowledgements}
We are grateful to David Basin and the Information Security Group at ETH Z{\"u}rich for hosting the six-month stay in Switzerland, during part of which we did this work. And thanks particularly to Andreas Lochbihler, who shared with us the probabilistic termination problem that led to it.

\bibliographystyle{plain}

\input{Termination.bbl}

\newpage\appendix
\Section{--- Sketch of Foster's proof \cite{Foster:1952aa}}\label{a1038}
We use the notation and definitions from \Sec{s1730} to present Foster's Theorem 2.

Recall that we have assumed that $S_0{=}\{s_0\}$, i.e.\ that termination occurs in a single state, and that we have adjusted (the assumed deterministic) $T$ so that it takes $s_0$ to itself.

Write $f^{(t)}_i$ for the probability that $T$ started from $s_i$ reaches $s_0$ for the first time in the $t$-th step and (as Foster does) write $p_{ij}$ for $T(i)(j)$, the probability that $T$ takes one step from $s_i$ to $s_j$; more generally write $p^{(t)}_{ij}$ for the probability that $T$ takes $t$ steps to do that. Foster remarks just after (F9) that a simple special case is where time-to-termination is bounded, but notes that such an assumption excludes the symmetric random walk and moves immediately to the more general case.
\footnote{\cite{Fioriti:2015} also treats the bounded-termination case explicitly.}

For the more general case we note first that for $i{>}0$ we have $f^{(t+1)}_i=\sum_j p_{ij}f^{(t)}_j$.
So if we were hopefully to proceed simply by setting $V(s_0){=}0$ and $V(s_i) = \sum_{1\leq t} f^{(t)}_i$ for $i{>}0$, then in the latter case we would check the super-martingale property (F1) by calculating
\begin{Reason}
\Step{}{\sum_j p_{ij}V(s_j)}
\Step{$=$}{\sum_j p_{ij}\sum_{1\leq t} f^{(t)}_j}
\Step{$=$}{\sum_{1\leq t}\sum_j p_{ij} f^{(t)}_j}
\StepR{$=$}{above and $i{>}0$}{\sum_{1\leq t}f^{(t+1)}_i}
\StepR{$\leq$}{(actually equal unless $f^{(1)}_i{>}0$)}{
 \sum_{1\leq t}f^{(t)}_i ,
}
\Step{$=$}{V(s_i)~,}
\end{Reason}
so that $V$ would in fact be an exact martingale in most of $S_\ast$.
\footnote{Think of the symmetric random walk, where everywhere-1 is an exact martingale except when $|x|{=}1$, where it is a proper super-martingale.}
But this looks too good to be true, and indeed it is: in fact $ \sum_{1\leq t} f^{(t)}_i = 1$ by assumption, so this is just the special case where $V$ is 1 everywhere except at $s_0$; and the martingale property is exact everywhere, except at states one step away from $s_0$. And this trivial $V$ does not satisfy \Progress.
\footnote{It is trivial in Blackwell's sense \cite{Blackwell:55}, a constant solution.}

Still, the above is the seed of a good idea. Using ``a theorem of Dini'' \cite[Foster's citation (4)]{Knopp:1928aa},
\footnote{There seems to be a typographical error here in Foster's paper, where he writes $\sum_{r=1}^\infty\lambda^{(r)}f^{(r)}_i$ instead of $\sum_{r=1}^\infty\lambda^{(r)}f^{(r)}_1$.}
that
\begin{quote}\it
If $c_n$ is a sequence of positive terms with $\sum_n{c_n}<\infty$, then also
\[
 \sum_n \frac{c_n}{(c_n{+}c_{n+1}+\cdots)^\alpha} \Wide{<} \infty
\]
when $\alpha{<}1$,
\end{quote}
Foster \emph{increases} the $f^{(t)}_i$ terms above by dividing them by {\scriptsize$\sqrt{f^{(t)}_1+f^{(t+1)}_1+\cdots}$}\,, which is non-zero but no more than one,
\footnote{It is the square-root of the probability that $s_1$ does not reach $s_0$ in fewer than $t$ steps.}
and still (as we will see) the new, larger terms still have a finite sum. (A minor detail is that he must show that the sum $f^{(t)}_1+f^{(t+1)}_1+\cdots$ does not become zero at some large $t$ and make terms from then on infinite: his assumption (F7) prevents that by ensuring that from no state in $S_\ast$ does a single $T$-step go entirely into $S_0$.) With the revised $V$ replacing the earlier ``hopeful'' definition, the calculation above becomes instead
\begin{Reason}
\Step{}{\sum_j p_{ij}V(s_j)}
\StepR{$=$}{revised definition of $V$, \\ and $V(s_0){=}0$}{
 \sum_{j\geq1} p_{ij}\sum_{1\leq t}\NF{f^{(t)}_j}{\sqrt{f^{(t)}_1+f^{(t+1)}_1+\cdots}}
}
\Step{$=$}{
 \sum_{1\leq t}\sum_j p_{ij} \NF{f^{(t)}_j}{\sqrt{f^{(t)}_1+f^{(t+1)}_1+\cdots}}
}
\Step{$=$}{
 \sum_{1\leq t}\NF{f^{(t+1)}_j}{\sqrt{f^{(t)}_1+f^{(t+`)}_1+\cdots}}
}
\StepR{$=$}{denominator is not increased}{
 \sum_{1\leq t}\NF{f^{(t+1)}_j}{\sqrt{f^{(t+1)}_1+f^{(t+2)}_1+\cdots}}
}
\Step{$\leq$}{
 \sum_{1\leq t}\NF{f^{(t)}_j}{\sqrt{f^{(t)}_1+f^{(t+1)}_1+\cdots}}
}
\Step{$=$}{V(s_i)~.}
\end{Reason}
This is encouraging: but we still must prove (F3) for our revised definition
\footnote{Note the $f$'s in the denominator are subscripted ``1'', not ``$i$''.}
\begin{Equation}\label{e0838}
 V(s_i)\Wide{=}\sum_{1\leq t}\frac{f^{(t)}_i}{\sqrt{f^{(t)}_1+f^{(t+1)}_1+\cdots}}~,
 \mbox{\quad for $i{\geq}1$}
\end{Equation}%
i.e.\ that it's finite for all $i$ and not only for the $i{=}1$ that Dini gave us; and we must show that is satisfies (F2), i.e.\ that it tends to infinity as $i$ does.

For the first, Foster proves that $V(s_i){\leq}V(s_1)/p^{(t')}_{1i}$ for any $i{>}1$ and some $t'{>}0$ with $p^{(t')}_{1i}{>}0$, which is one place he uses (F7), in particular that every $s_i$ is accessible from $s_1$.
\par Specifically, he reasons as follows:
\begin{enumerate}[(i)]
\item For that $t'$ and any $t$ we have $f^{(t'+t)}_1\geq p^{(t')}_{1i}f^{(t)}_i$, because we know that $s_1$'s journey to $s_0$ can go via $s_i$. 
\item The numerator $f_i^{(t)}$ in \Eqn{e0838} can therefore be replaced by $f^{(t'+t)}_1/p^{(t')}_{1i}$ provided $(\leq)$ replaces the equality.
\item The sum in the denominator of \Eqn{e0838} can be adjusted to start at $t'{+}t$ rather than $t$, still preserving the inequality.
\item The overall sum in \Eqn{e0838} of non-negative terms for $V(s_i)$ is now the ``drop the first $t'$ terms suffix'' of that same sum for $V(s_1)$, which we already know to be finite (from Dini), but divided by $p^{(t')}_{1i}$ which we know to be non-zero.
\end{enumerate}

For the second, Foster uses the $\delta$ from (F8), showing that $V(s_i)$ is at least $\NF{(1{-}\delta)}{\sqrt{f^{(t_i)}_1+f^{(t_i+1)}_1+\cdots}}$ where $t_i$ is the number of steps after which $s_i$ reaches $s_0$ with probability at least $\delta$ for the first time. By (F8) that $t_i$ tends to infinity as $i$ does, and thus so does $V(s_i)$.
\par His detailed reasoning is as follows:
\begin{enumerate}[(i)]
\item Since $t_i$'s tending to infinity is all that is required, any at-most-finite number of $i$'s where $t_i{=}0$ can be ignored. Thus pick $t_i{\geq}1$.
\Cf{I'm not sure why $t_i{\geq}1$ helps, though.}
\item Then $V(s_i)$ is at least $\sum_{1{+}t_i\leq t}\NF{f^{(t)}_i}{\sqrt{f^{(t)}_1+f^{(t+1)}_1+\cdots}}$\,, a suffix of its defining series \Eqn{e0838}.
\item Since the denominators only decrease, we can replace all of the denominators by {\scriptsize${\sqrt{f^{(t_i)}_1+f^{(t_i+1)}_1+\cdots}}$}\, while making the sum only smaller.
\item From (F8) however and the choice of $t_i$ we know that $\sum_{t_i\leq t}f_0^{(t)}$ is no more than $1{-}\delta$. Thus similarly we can replace $f^{(t)}_i$ by $1{-}\delta$ and remove the summation.
\item We are left with $V(s_i)\geq \NF{(1{-}\delta)}{\sqrt{f^{(t_i)}_1+f^{(t_i+1)}_1+\cdots}}$\,, as appealed to above.
\end{enumerate}

\bigskip That completes the proof sketch.

\newpage\Section{Loop rule pr{\'e}cis (from \Sec{ss0154})}\label{s1728}
To be clear, we interpret here our Lem.\,7.3.1 from \cite{McIver:05a}, which includes demonic nondeterminism. The lemma reads
\begin{quote}
Let invariant $I$ satisfy $[G ]{\ast}I \RRightarrow \WLP.\textit{body}.I$. Then $I\&T \RRightarrow \WP.\textit{loop}.([\overline{G}]{\ast}I)$, where $T$ is the termination probability $\WP.\textit{loop}.[\textsf{true}]$ of the loop.
\end{quote}
We note that
\begin{itemize}
\item $G$ is a predicate over the variables of the program.
\item Square brackets $[\cdot]$ convert Boolean \True,\False\ to numeric 1,0 respectively.
\item $I,T$ are real-valued expressions over the variables of the program, in the interval $[0,1]$.
\item \WP\ and \WLP\ are the probabilistic generalisations of Dijkstra's weakest- and weakest-liberal precondition functions respectively \cite{Dijkstra:76,Kozen:85,Morgan:96d,McIver:05a}.
\item[]
\item $G$ is the loop guard, for us a predicate characterising $S_\ast$.
\item $I$ is the loop invariant, for us (confusingly) the variant $V$.
\item $\RRightarrow$ is the $\leq$ relation on functions, defined pointwise (as usual).
\item $[G ]{\ast}I \RRightarrow \textrm{wlp}.\textit{body}.I$ says that the expected value of $I$ (our $V$) after one transition is at least as great as its actual value at the state from which the transition was taken. (The $[G]{\ast}$ means that the relation is mandated only when $G$ holds, i.e.\ within $S_\ast$.) Thus it is this condition that states that $V$ is a sub-martingale on $S_\ast$.
\item In general $A\&B$ is $A+B-1\Max0$ when $0{\leq}A,B{\leq}1$. Thus when $T{=}1$ we have that $I\&T = I$.
\item $\overline{G}$ is $G$'s negation, so that $[\overline{G}]{\ast}I$ is our $V$ set to zero on $S_\ast$, equivalently our $V$ restricted to $S_0$.
\item[]
\item The lemma's inequality thus states
\begin{quote}
If termination occurs with probability one ($T{=}1$ everywhere), then the value of $V$ on the starting state is no more than ($I\&T\RRightarrow$) the expected value of $V$ on $S_0$ when escape from $S_\ast$ has occurred ($\textrm{wp}.\textit{loop}.([\overline{G}]{\ast}I)$).
\end{quote}
\end{itemize}
\end{document} 

\newpage\MakeGreenRoom
\newpage\MakeEndNotes
\newpage\MakeSargasso
\Section{Taken from \Sec{s1958}}
\Subsection{Two-dimensional random walk on diamonds}

I'll write this as a program, but there is a nice picture that you can draw of diamonds.
\Cf{\A{Annabelle}, are you sure this program doesn't have some typo's? I can't quite figure it out.}
\Af{I think this example should be removed.}
\begin{Prog}
int x, y 
while (x != 0 and y != 0) {
   if (x=0 or y=0) then {
     if (x=0) then y:= |y|-1 [1/2] -|y|+1;
                         Skip [1/2]  (x:= x+1 [1/2] y:= x-1);
     if (y=0) then x:= |x|-1 [1/2] -|x|+1; 
                         Skip [1/2] (y:= y+1 [1/2] y:= y-1);
   }
  else {
     x:= x+1 [1/2] x:= x-1;
     y:= y+1 [1/2] y:= y-1
   }  
}
\end{Prog} 

Variant is $|x| + |y|$.

This is the ordinary random walk where we imagine that the particle at $(x,y)$ (normally) moves diagonally NE, SE, NW, SW. On the boundaries ($x=0$ and $y=0$) it first moves one step towards the origin before making its diagonal hop.  The encoding of this was annoying for the cases where either $x=0$ or $y=0$, so  I put a fifty/fifty hop between moving one place in, or the reflection of that move. 
Overall this is a particle that can move freely around larger and larger diamonds, or hop out to the next larger one, or back to the next smaller one. 

It probably looks better on a diagram --- however the encoding is a little tricky -- if one is not careful then one of the conditions breaks. Does this make a point?

\newpage\Section{\Ax A conjectured variant \\ for the two-dimensional random walk}

Take the ``standard'' two-dimensional random walker, and use the following variant in which we write $\Lgg(x)$ for $\log(\log(x))$:

\[
\begin{array}{ll@{\hspace{3ex}}l@{\hspace{6ex}}p{20ex}}
V(x,y) &=& \Lgg (x^2 + y^2) & if $x^2{+}y^2 > 1$ \\
          & =& 0                        &     otherwise.
\end{array}
\]

This variant is based on the Euclidean distance to the origin, and it is clear that there is always a $1/4$ probability of strictly decreasing $V$. Thus we only need show strict decrease of the variant, i.e.\ we must show:
\[
 \begin{array}{ccl}
       && \Lgg ((x{+}1)^2 + y^2) \\
   &+  & \Lgg ((x{{-}}1)^2 + y^2) \\
   &+  & \Lgg (x^2 + (y{+}1)^2) \\
   &+ & \Lgg (x^2 + (y{{-}}1)^2) \\[1ex]
   < && 4\Lgg (x^2{+}y^2)~,
\end{array}
\]
assuming that $x^2{+}y^2 > 1$ initially.  There are two cases. The first is if both $x, y \neq 0$. The second is if one of $x, y$ is zero.

\underline{Case: either $x$ or $y$ is zero.}

First we show that for $V'(x,y) = \log(x^2 {+} y^2)$, the average reduction is strict.  Then we observe that for any variant for which the average reduction is strict, then the $\log$ of the variant also satisfies strict decrease of the average move. 

Suppose wlog that $x=0$. Then we want to show

\[
 1/2\log (1{+}y^2) + 1/4 \log(y{+}1)^2 + 1/4\log(y{-}1)^2 < \log(y^2)~.
\] 

Applying some log rules:
\[
\log (1{+}y^2) +  \log(y{+}1) + \log(y{-}1) \Wide{<} \log(y^4)~.
\] 

Another log rule:
\[
\log ((1{+}y^2)(y{+}1)(y{-}1)) \Wide{<} \log(y^4)~.
\] 

But notice that $((1{+}y^2)(y{+}1)(y{-}1))  = y^4 {-}1$, allowing us to simplify the lhs (above), so finally we are done if 
\[
\log (y^4{{-}}1) \Wide{<} \log(y^4)~,
\] 
which is true by monotonicity of $\log$. Hence since $V'$ has the property we want, so too does $\log V'(x, y) = \Lgg (x, y)$.

Unfortunately I couldn't get the other case to work. Sorry! Suggest using e.g.\  Taylor expansions of $\Lgg(x^2{{+}}y^2)$.

\newpage\Section{Termination cannot be everywhere almost-certain \\ for bounded martingales}\label{ss0154}
Using the techniques here, and ideas from \cite{Chatterjee:16}, we can extend Blackwell's \Thm{t1217} to a more general situation.

\begin{Theorem}{Non-termination}{t0804}
Let $T$ be a transition system as above, and let $V$ be a variant satisfying the following conditions:
\begin{itemize}
\item For all $s$ in $S_\ast$ and $\delta$ in $T(s)$ we have $\Exp_{\delta}V{=}V(s)$, i.e.\ that $V$ is an exact martingale, neither super- nor sub-.
\item $V$ is non-constant and bounded on $S_\ast$.
\end{itemize}
(Note that we do not have any progress requirements.)

Then there is at least one $s$ in $S_\ast$ from which the probability of termination under $T$ can be is strictly less than $1$.
\footnote{
Note that our own termination rule  is not able to prove almost-sure termination with only these conditions for $V$, as discussed above in \Sec{ss1958}. We would need additionally some version of \Progress, either $p,d$ or $\nabla$.

This theorem however has more significance: it shows not only that we can not apply our rules, but that in fact everywhere-terminating is \emph{refuted} under these conditions. That is, under these conditions there \emph{cannot be} suitable $p,d$ or $\nabla$.
} 

\Proof
Let $B$ be the least upper bound of $V$, and choose some $s$ with $V(s){<}B$, which must be possible since otherwise $V(s)$ would be $B$ for all $s$, i.e.\ it would be constant.
Create the process $T'$ which mimics $T$ except that it  terminates not only if it reaches $S_0$ but also if it reaches any $s'$ where $V(s'){>}V(s)$.  Define variant $V'(s')$ over this new process so that it is equal to $V(s')$ when $V(s'){\leq}V(s)$ but is equal to $B$ when $V(s'){>}V(s)$. Note that this new $V'$ is a sub-martingale on $\{s'{\In}S_\ast\mid0{<}V(s'){\leq}V(s)\}$, since $B$ the least upper bound of $V$ on $S_\ast$.

Now assume for a contradiction that termination under $T$, i.e.\ reaching $S_0$, is \AC\ from $s$; it will then be \AC\ in $T'$ as well, since the termination conditions for $T'$ are more liberal than those for $T$. As in the proof of \Thm{t1356}, let $\delta$ be one of $T'$'s has-terminated distributions.
\footnote{There might be more than one if there is demonic nondeterminism.}
Let $p$ be the probability that $\delta$ assigns to $S_0$, so that $\delta$ assigns the  complementary probability $1{-}p$ is to termination in a state $s'$ where $V'(s'){=}B$. We now reason
\begin{Reason}
\StepR{}{$V'$ is a sub-martingale under $T'$, \\ and $T'$ starts from $s$}{
 V'(s)~\leq~p{\times}0+(1{-}p)B
}
\Space
\StepR{whence}{$V'(s){=}V(s)$}{
 p~\leq~1{-}V'(s)/B~=~1{-}V(s)/B~.
}
\end{Reason}
Since $s$ is in $S_\ast$ we have $V(s){>}0$, thus also $p{<}1$ which contradicts our assumption that $T$, and thus $T'$ was terminating from $s$.
\Cf{\Label{n0924}But this $p$ applies only to $V{=}0$ states that are reached without going through a $V{>}V(s)$ -state first --- that is, there could be other paths from $s$ to $S_0$ that this $p$ does not take into account. And I think this issue applies to the earlier ``other way up, with $\varepsilon$'' version \A{\Sec{Sargasso-ss0154}},\Sec{Sargasso-ss0154a} as well.

But I do not think it affects our termination rule (although the same effect occurs) since there we are bounding $z$ (here called $p$) from below: it does not matter if we missed some paths to $V{=}0$, because including them only makes the probability of termination greater still.

But do we need Azuma's inequality here \cite{Chatterjee:16}?

See \Fn{n0925}.}
\end{Theorem}


Blackwell's \Thm{t1217} is a special case of the above where the process is a random walk  in one dimension. Note that we are  essentially using $V$  as a ``refuting martingale'' as in \cite{Chatterjee:16}.

\newpage
\Section{Alternative to {\Progress}}\label{s2117}
Here is an alternative reformulation of {\Progress}:
\begin{description}
\item[$V$ must make progress towards zero ---] There is a non-increasing function $H$ from the non-negative reals to itself such that for every state $s$ in $S_\ast$ there is a non-zero probability that $V(s)$ will be decreased by at least $H(V(s))$. That is,
\begin{quote}
 For any $s$ in $S_\ast$ and $\delta$ in $T(s)$, there must be an $s'$ in the support of $\delta$ such that $V(s')\leq v{-}H(v)$, where $v{=}V(s)$.
\end{quote}
\end{description}
The first part of the soundness argument of \Sec{s1351} would be altered to read
\begin{quote}
Fix some $v{>}0$. From {\Progress} ($V$ must make progress towards zero), for any $s$ with $V(s){\in}(0,v]$ there must be a non-zero probability of decrease of $V$ by at least $H(v)$, where that $H(v)$-minimum applies uniformly to all such $s$ because function $H$ is non-increasing. That means that from every such $s$ there is a escaping path of length at most $v/H(v)$, i.e.\ a finite path, which thus has non-zero overall probability. Thus, by the zero-one Law, escape from $(0,v]$ is \AC.
\Cf{Refer Hart et al.\ and ourselves.}
It is possible however that the escape occurs not by setting $V$ to 0 but rather by setting $V$ to some value greater than $v$. 
\end{quote}
The rest of the argument proceeds as before, replacing constant $H$ by $v$. That gives
\begin{quote}
Suppose now that for this $s$ the probability of eventually escaping at 0 is $z$ and for eventually escaping above $v$ is $h$. Because escape is \AC, we have $z{+}h=1$.
\Cf{With non-determinism that would read
\begin{quote}
Take any distribution $\delta$ representing $s$'s escape from $(0,v]$; let $z$ be the aggregate probability it assigns to states $s'$ with $V(s'){=}0$; let $h{=}1{-}z$ be the remaining probability, i.e.\ that it assigns to states $s''$ with $V(s''){>}v$.
\end{quote}}
The expected value of $V$ over that has-escaped distribution is at least $z{\times}0+h{\times}v$, since the actual value of $V$ in the $(h{\times})$-case is at least $v$. But by {\SMart} ($V$ is a super-martingale wrt.\ $T$), we know that the expected value of $V$ when escape from $(0,v]$ occurs, having started from $s$, cannot be more than $V(s)$ itself. So we have $hv\leq V(s)$, whence $z=1{-}h\geq1{-}\NF{V(s)}{v}$.

To conclude, we note that the inequality $z\geq1{-}\NF{V(s)}{v}$ just above holds for any choice of $s,v$ with $V(s){\leq}v$ and in particular, having fixed an $s$, we can make $v$ arbitrarily large. Thus $z$ must be 1 for all $s$.
\end{quote}

The argument from \Sec{s0819} that our rule subsumes Fioriti and Hermanns' also becomes much simpler. If $\Exp_\delta V$ is at least $\varepsilon$ below $V(s)$ for any $\delta$ in $T(s)$,
\Cf{We are using the nondeterministic formulation here.}
which is Fioriti and Hermanns' condition, then for some $s'$ in the support of $\delta$ we must have $V(s')\leq V(s){-}\varepsilon$. Thus we can define our $H$ to be the constant function everywhere $\varepsilon$.

\newpage
\Subsubsection{Are there alternatives formulations of {\Progress}?}
Yes. The proof of the rule (\Sec{s1351}) uses {\Progress} only to show bold escape from an arbitrary but bounded variant-region $(0,H]$, that is that for any $0{<}V{\leq}H$ the probability that eventually either $V{=}0$ or $V{>}H$ is bounded away from 0, where the bound can depend on $H$. (It is {\SMart} that then converts that to \AC\ escape to $V{=}0$, without any $H$.) Any other condition with the same force would do. Examples are
\begin{enumerate}[(i)]
\item $V$ strictly decreases with some non-zero probability (but not necessarily boldly), and has no accumulation points.
\item As just above, but replacing ``no accumulation points'' with ``for any $H$ there are only finitely many $V$-values in $[0,H]$.''
\item Like Fioriti and Hermanns' condition, but weakened so that the boldness of the super-martingale's decrease can ``depend on $H$'' rather than having to be a constant $\varepsilon$. It would then read
\begin{quote}
There must be a non-increasing function $E$ from the non-negative reals to the non-negative reals, and $V$ as before, such that from every state $s$ and $\delta$ in $T(s)$ we have $\Exp_\delta V \leq V(s){-}E(V(s))$.
\end{quote}
An interesting advantage of this one is that it makes \Progress\ unnecessary, so that we have just once condition. See \Sec{s0807} for a discussion of this alternative.
\item $V$ decreases at least some $e_H$ with non-zero probability.
\Cf{This alternative just occurred to me: it's a strict weakening of what we have, because it imposes no $p_H$. But do we need that $p_H$ actually? Consider: For any $s$ with $0{<}V(s){\leq}H$ there is some $p_s{>}0$ such that $V(s)$ decreases at least $e_H$. It can decrease that way only $V(s)/e_H$-many times, however, a finite number, and so we have no Zeno issues with decreasing $p_s{>}p_{s'}{>}p_{s''}{>}\cdots$ etc. See \Sec{Sargasso-s2117}. \Ax Don't think so (says Annabelle); you have forgotten that there can be a lot of Zeno hidden in each value of $V$ separately. Sorry.}
\end{enumerate}
The reason we don't choose one of those (at the moment) is that a subsidiary aim is to have a rule that clearly generalises Fioriti and Hermanns' (\Sec{s0819}).

\Subsubsection{\Cx Can you think of any more? \ATD}~\\
\Ct{What should be here is pre-emptive answers to a reader's sanity-check questions of himself: ``This is ridiculous: it can't work because of\ldots''. Let's anticipate those questions, and answer them here.}

\newpage
\begin{description}
\item[$V$ must make progress towards zero --- ] For any non-negative real $H$ (for ``high'') there are two reals $e_H{>}0$ and $0{<}p_H{\leq}1$, possibly depending on $H$, such that $V$ itself (i.e.\ not just its expected value) is guaranteed with probability at least $p_H$ to decrease by at least $e_H$ whenever a transition is taken from a state in $S_\ast$ where $V$ is no more than $H$ in the first place.
\Cf{I think we need an ``or set V to zero'' possibility here: otherwise the condition can't be satisfied if $V$ is close to 0 already.} 
That is,
\begin{quote}
 For any $H{\geq}0$ there are reals $e_H{>}0$ and $0{<}p_H{\leq}1$ such that for any state $s_\ast$ in $S_\ast$ with $V(s_\ast)\leq H$ we have
 \[
  T(s_\ast)_{\{s'\mid V(s')\leq V(s_\ast){-}e_h\}} ~\geq~ p_H\quad.
 \]
\end{quote}
\Cf{We ``abuse notation'' by writing $\delta_{S'}$ for $\sum_{s'\In S'} \delta_{s'}$~.}
Our innovation here is to impose the usual ``make progress'' criterion not on $S_\ast$ as a whole, but instead independently on successively larger portions of it.
\end{description}

\newpage
\Subsection{Proof of soundness}
\begin{Theorem}{Soundness of \Sec{s1203}}{t1356}
The ``proof of almost certain termination technique'' described in \Sec{s1203} is sound.
\Proof
Fix some $H{\geq}0$. From {\Progress} ($V$ must make progress towards zero), there must be $e,p$ for this $H$ (having fixed $H$ we can drop the subscripts) such that every transition from any state $s$ with $V(s){\leq}H$ is guaranteed with probability at least $p$ to decrease $V$ by at least $e$. Fix such an $s$, i.e.\  with $V(s){\leq}H$. The probability that $V$ will eventually become 0 via transitions from that $s$ is no less than $p^{H/e}$, since taking the probability-at-least-$p$ option to ``decrease $V$ by at least $e$'' suffices if it is taken at least $H/e$ times in a row.

Since this applies for all $s$ with $V(s){\leq}H$, and $p^{H/e}$ is independent of $s$, the probability of transitions' leaving subset $\{s\mid V(s){\leq}H\}$ eventually is bounded away from zero for that subset of $S$. By the zero-one Law, that termination probability must therefore be one.
\Cf{Refer Hart et al.\ and ourselves.}

It is possible however that the escape occurs not by setting $V$ to 0 but rather by setting $V$ to some value greater than $H$. Suppose the probability of the former is $z$ and for the latter is $h$ (for this $s$). And $z{+}h=1$ since escape one way or the other is \AC.
\Cf{Will need to explain how the nondeterminism is working here.}
The expected value of $V$ over that distribution is at least $z{\times}0+h{\times}H$, since the actual value of $V$ in the latter case is at least $H$.

But by {\SMart} ($V$ is a super-martingale wrt.\ $T$), we know that the expected value of $V$ when escape occurs, having started from $s$, cannot be more than $V(s)$ itself. So we have $hH{\leq}V(s)$, whence $z=1{-}h\geq1{-}\NF{V(s)}{H}$.

To conclude, we note that the inequality $z\geq1{-}\NF{V(s)}{H}$ just above holds for any choice of $s,H$ and, in particular, having fixed an $s$ we can make $H$ arbitrarily large. Thus $z$ must be 1 for all $s$.
\end{Theorem}

\newpage
\Section{Proof that our rule subsumes Fioriti and Hermanns'}\label{s0819}
Fioriti and Hermanns do not have our {\Progress} condition; instead he requires bounded-away-from-zero decrease of the variant (where we require only that it not increase).

Suppose therefore that we have a Fioriti and Hermanns -style variant, i.e.\ such that for all $s$ in $S_\ast$ we have
\begin{Equation}\label{e1441}
 \Exp_{T(s)}V \Wide{\leq} V(s){-}\varepsilon
\end{Equation}%
for some $\varepsilon{>}0$ fixed across the whole system; this clearly satisfies our {\SMart} requirement. We now show that it satisfies {\Progress} as well.

Fix $H,s$ with $V(s){\leq}H$, and resolve nondeterminism to choose some $\delta$ in $T(s)$.
\Cf{Now $T(s)$ is a set; this is how we'll explain nondeterminism, when we get around to it.}
We have from \Eqn{e1441} that $\Exp_\delta V \leq V(s){-}\varepsilon.$

Set $e_H$ to $\NF{\varepsilon}{2}$, and let $p$ be $\delta_{[0,V(s){-}e_H]}$, i.e.\ the probability that $\delta$ decreases $V$ by at least $e_H$. The greatest decrease in the $\delta$-expected value of $V$ cannot be more than the decrease that would accrue if $\delta$ placed all that weight $p$ at 0 and the remaining weight $1{-}p$ at $V(s){-}e_H$:
\Cf{Actually, just a smidgeon more than that.}
that is, the greatest decrease in the expected value of $V$ cannot be more than
\[
 p{\times}H + (1{-}p){\times}e_H~,
\]
since we know $V(s){\leq}H$. But from Fioriti and Hermanns' condition \Eqn{e1441} that decrease must be at least $\varepsilon$, and so we reason
\begin{Reason}
\Step{}{p{\times}H+(1{-}p){\times}e_H ~\geq~ \varepsilon}
\Step{iff}{p{\times}H+e_H-p{\times}e_H ~\geq~ \varepsilon}
\StepR{iff}{$e_H=\varepsilon/2$}{p{\times}(H{-}e_H) ~\geq~ e_H}
\Step{iff}{p ~\geq~ e_H/(H{-}e_H)~.}
\end{Reason}
Thus it's sufficient for {\Progress} to set $e_H$ to $\varepsilon/2$ and $p_H$ to $\varepsilon/(2H{-}\varepsilon)$. We note  that $e_H$ is fixed, independent of $H$ (but dependent on Fioriti and Hermanns' global $\varepsilon$); on the other hand $p_H$ decreases as $H$ increases, indeed can become arbitrary small. But for each $H$ separately it is greater that zero, which is all that {\Progress} requires.

Since Fioriti and Hermanns' rule is complete for systems with finite expected time to termination, the result above means our proposed rule is also complete for that class. But --as observed in \Sec{s1203}-- our rule also applies to the unbounded random walk, where the termination time is infinite.

\newpage
\Section{Alternative to {\Progress}}\label{s0807}
First --a sanity check-- can we still do with this formulation what Fioriti and Hermanns cannot? The unbounded symmetric random walk with variant being distance from 0 won't work any more: i.e.\ taking $S$ to be the integers and $V(s)\Defs s$ makes $V$'s expected value a constant. So, instead, define $V(s)\Defs \sum_{1\leq n\leq s} 1/s$ or similar: any non-decreasing and unbounded concave function $f$ should do. (e.g.\ $\log$).
\Cf{Actually, I think using $1{+}\log$ for the walk would be cool. We don't require it to be defined at 0.}
Then define $\nabla(s)\Defs 1/s - (\NF{1}{s{-}1}+\NF{1}{s{+}1})/2$, or more generally $f(s) - \NF{f(s+1)+f(s-1)}{2}$. And that's it.

The reason that doesn't work for Fioriti and Hermanns is that $\nabla(s)$ is not bounded away from 0.

What needs proof is that the revised condition guarantees eventual escape from any variant-region $(0,H]$, and the proof of that we have already done elsewhere (in this draft) for other purposes. I will copy (and specialise) it here.
\Cf{But we could instead just appeal to Fioriti and Hermanns' result, specialised to the case where the variant is bounded. The proofs below could then be in an appendix.}

\begin{Lemma}{Decrease of expectation}{l0835}
Let $f$ be a non-negative function over the non-negative reals, and let $y,y'$ be non-negative reals; let $\delta$ be a discrete distribution on the non-negative reals. Then
\begin{Equation}\label{e0955}
 \Exp_\delta f \leq y \WideRm{implies} \delta_{\{x\mid f(x){<}y'\}}\geq1{-}\NF{y}{y'}~.
\end{Equation}%

\Proof
Let $p$ be the aggregate probability that $\delta$ assigns to $\{x\mid f(x){\geq}y'\}$. Then the smallest possible value of $\Exp_\delta f$  is $py'$, occurring when $f(x){=}0$ for all $x$ with $f(x){<}y'$ and $f(x){=}y'$ for all $x$ with $f(x){\geq}y'$. Thus $py'\leq\Exp_\delta f \leq y$, whence $p\leq\NF{y}{y'}$ and so $ \delta_{\{x\mid f(x){<}y'\}}=1{-}p\geq1{-}\NF{y}{y'}$.
\end{Lemma}

\begin{Lemma}{Guaranteed decrease of variant}{l0926}
Let $V,S,T$ etc.\ be as above, and let $\nabla$ be a non-increasing function on the non-negative reals. Suppose for all states $s$ in $S_\ast$ with $0{<}V(s){\leq}H$ we have that any $T$-transition is guaranteed to decrease the expected value of $V$ by at least $\nabla(V(s))$.

Then any transition $T$ is guaranteed with probability $p{>}0$ to decrease the \emph{actual} value of $V$ by at least $e{>}0$, where $e\Defs\NF{\nabla(H)}{2}$ and $p\Defs\NF{e}{H{-}e}$.

\Proof
Since $\nabla$ is non-increasing, in fact any $T$-transition is guaranteed to decrease the expected value of $V$ by at least $\nabla(H)\leq \nabla(V(s))$, the advantage here being that $\nabla(H)$ is independent of $s$.

Let $\delta$ in $T(s)$ be any $T$-transition from $s$, and for \Lem{l0835} set $y=V(s){-}\nabla(H)$ and $y' = V(s){-}\NF{\nabla(H)}{2}$ and $f{=}V$. Then $\delta$ is guaranteed with probability at least $1{-}\NF{y}{y'} = 1-\frac{V(s){-}\nabla(H)}{V(s){-}\NF{\nabla(H)}{2}}$ to decrease $V$ by at least $V(s){-}y'=\NF{\nabla(H)}{2}$.

Thus indeed $e=\NF{\nabla(H)}{2}$. Because $V(s){\leq}H$, the probability is in turn at least $1{-}\frac{H{-}\nabla(H)}{H{-}\NF{\nabla(H)}{2}} = \NF{e}{H{-}e}$, which is $p$.
\end{Lemma}

\begin{Lemma}{Eventual escape from $(0,H]$}{l1010}
Under the conditions of \Lem{l0926}, escape from the region $\{s\mid0{<}V(s){\leq}H\}$ is \AC.

\Proof We use the terminology of \Lem{l0926}.

Since every $s$ in that region is guaranteed with probability at least $p{>}0$ to decrease $V(s)$ by at least $e{>}0$, with $p,e$ independent of $s$, every $s$ has probability at least $p^{H/e}$ of reaching eventually an $s'$ with $V(s'){=}0$, thus eventually escaping the region with probability bounded away from 0.

By the zero-one Law, 
\Cf{Refer Hart et al.\ and ourselves.}
escape from the region is in fact \AC. (Note however that the escape might occur to an $s'$ with $V(s'){>}H$.)
\end{Lemma}

\newpage
\Section{Informal proof of the new rule: first attempt\protect{\Cf{A direct proof is given in \Sec{s1351}}}}\label{s1352}

We begin with a lemma concerning bounded random walks.
\begin{Lemma}{Bounded random walk escape}{l1030}
Consider a possibly demonic bounded random walker in $(0,N)$ that begins at some $0{<}n{<}N$. It can jump to any position in $[0,N]$ provided that its probability of moving up is bounded away from zero at each position $n$ separately,
\footnote{The bounding away from zero here is over the possible nondeterminism at each position separately. But because there are only finitely many positions $n$ in $(0,N)$, the separate bounds away from zero are equivalent to a uniform bound over the whole of $(0,N)$. But later the distinction will become important.}
and that the least expected value of its new position is no less than its current position.
\footnote{We say \emph{least} expected value because of the possible nondeterminism.}
Under the conditions above, the probability of the walker's reaching $N$ is at least $n/N$.
\footnote{This is a generalisation of the usual symmetric random walk, where the probability of reaching $N$ from starting position $n$ is exactly $\NF{n}{N}$.} 
\Proof We have assumed that from every position in $(0,N)$ there is a non-zero probability that the walker will move up. (When we consider nondeterminism, we will use the fact that the nondeterminism is finite to argue that the probability of the walker's moving up is bounded away from zero throughout all of $(0,N)$.) Thus with probability 1 the walker must reach either 0 or $N$ eventually: it cannot remain forever in $(0,N)$.

Now the expected value of the walker's final position is at least $pN$, where $p$ is the least probability of its reaching $N$ rather than 0. Thus we have $pN>n$, because $n$ is the (expected) value of its initial position and that expected value can only increase. Thus  $p>n/N$.
\end{Lemma}

We now continue by adapting \Lem{l1030} to our current purposes, in particular a decreasing variant (rather than increasing position).
\begin{Corollary}{Extended lower bound}{c1100}
Under the same conditions as \Lem{l1030}, except that the lower bound is $-L$ for some integer $L{\geq}0$, the probability of reaching $N$ is at least $(n{+}L)/(N{+}L)$.
\Proof
Just shift the whole argument of \Lem{l1030} up by $L$.
\end{Corollary}

\begin{Corollary}{No lower bound}{c1103}
Under the same conditions as \Lem{l1030}, but without any lower bound at all (i.e.\ neither 0 nor the more general $-L$), the probability of reaching $N$ eventually is at least \NF{1}{2} from any starting point.
\Proof
Given starting point some $n$, choose $L$ so that $(n{+}L)/(N{+}L) \geq \NF{1}{2}$.
\Cf{ 
In detail:
\begin{Reason}
\Step{}{ (n{+}L)/(N{+}L) \geq \NF{1}{2}}
\Step{if}{2n{+}2L \geq N{+}L}
\Step{if}{L\geq N{-}2n~,}
\end{Reason}
so that there is always such an $L$. If $N{-}2n\leq0$, that is $n\geq N{/}2$ ``already'', then of course $L{=}0$ suffices.
} 
\end{Corollary}

We now move the walker down rather than up:
\begin{Corollary}{Almost-certain termination of random walker}{c1113}
Consider a generalised, possibly nondeterministic, random walk on the non-negative number line that on each step can jump from its current position $n$ to any non-negative $n'$ provided the expected value of its new position $n'$ is no more than its current position $n$ and that its probability of moving down is bounded away from zero at each position separately.

Then no matter where it starts it will reach 0 eventually with probability 1.
\Proof
By the zero-one Law
\Cf{Refer Hart et al.\ and ourselves.}
we can strengthen the result of \Cor{c1103} to conclude that the probability of reaching $N$ eventually is in fact 1, not only $\NF{1}{2}$. The result follows by swapping ``up'' and ``down'' and taking $N$ to be 0.
\end{Corollary}

\newpage
{
\Section{Boundedness is not possible for non-strict super-martingales}\label{ss0154a}

Blackwell's result seems to extend to a more general situation as follows.

Let $T$ be a transition system as above. Let $V$ be a variant. Suppose $V$ and $T$ satisfy the following conditions:

\begin{enumerate}
\item 
 \text{For all $s$ in $S_\ast$ and $\delta$ in $T(s)$ we have}\hspace{5ex} $\Exp_{\delta} V ~=~ V(s)$,

\item $V$ is non-constant and bounded.

\item The progress properties are the same as for the original rule.

With these conditions, the probability of termination is strictly less than $1$.

\end{enumerate}

\begin{proof}

Let $B$ be equal to the least upper bound of $V$. 
Observe that $B - V(s)$ is still a super-martingale, but decreases in value on states which require more steps to terminate. This is now similar to the repulsing super-martingale of Chatterjee et al. 
We show that for any $s'$ in $S_\ast$, the chance that it terminates is bounded away from $1$.

Given  any $s'$, choose $\varepsilon >0$ choose such that $B- V(s') > \varepsilon$. Create the process which follows $T$ except that if it  terminates if it reaches $S_\ast$ or any $s$ where $B-V(s) \leq \varepsilon$. 

Define $V'(s)$ to be $B-V(s)$ if $B-V(s) \geq \varepsilon$ and $0$ otherwise.

Let $p_\varepsilon$ be the chance that when the process starts at $s'$ it terminates at $S_\ast$. This process terminates almost surely for the reasons given \Thm{t1356}, so that with probability $1-p_\varepsilon$, the process terminates in a state where $V'(s) \leq \varepsilon$. But now we reason:

\begin{Reason}
\Step{}
{V'(s') \geq p_\varepsilon B + (1-p_\varepsilon)\times 0}
\Step{iff}
{V'(s')/B \geq p_\varepsilon }
\end{Reason}

Thus the chance that the original process terminates, is no more than $V'(s')/B = (B-V(s'))/B < 1$.
\end{proof}

If $V$ is unbounded, it is not possible to create a repulsing super-martingale. If the average decrease is strict, but bounded then $B-V$ is not a repulsing super-martingale. So, for example, if the probability of reaching $S_\ast$ is $1$ from every state, we could set $V(s)=1$ for all states outside $S_\ast$, but $1-V$ is not a repulsing super-martingale.
} 

\end{document}

%% file: MacrosGeneral.tex
\usepackage{amsfonts}			
\usepackage{amsmath}			
\usepackage{amssymb}
\usepackage{bbding}				
\usepackage{chapterbib}			
\usepackage{endnotes}			
\usepackage{enumerate}			
\usepackage{graphicx}			
\usepackage{hyperref}			
\usepackage{listings}			
\usepackage{longtable}			
\usepackage{stmaryrd}			
\usepackage{subfigure}			
\usepackage[normalem]{ulem}		
\usepackage{units}				
\usepackage{url}				
\usepackage{wasysym}			
\usepackage{wrapfig}
\usepackage{xcolor}				

\usepackage{latexsym}

\def\Vhrulefill{\leavevmode\leaders\hrule height 0.7ex depth \dimexpr0.4pt-0.7ex\hfill\kern0pt}

\newcommand\ImageInText[4]%
{\makebox[0pt][l]{%
{\def\Up{#1}\def\Right{#2}
\hfill%
\raisebox{\Up}[0pt][0pt]{
\makebox[0pt][l]{\hspace{\Right}
\includegraphics[scale=#3]{#4}%
}}}}}

\newcommand\ImageInTextBlock[7]%
{\def\Height{#1}
 \def\Depth{#2}
 \def\Width{#3}
 \def\Up{#4}
 \def\Right{#5}
\raisebox{\Up}[\Height][\Depth]{
\makebox[\Width][l]{\hspace{\Right}
\includegraphics[scale=#6]{#7}%
}}}

\newcommand\FirstLineOfTheorem[1][\empty] {\ifx#1\empty\quad\else~\hrulefill~#1\\\fi}
\newcommand\MakeFact[2]{\newenvironment{#1}[2]{\begin{#2}\label{##2}\textit{##1}\rm\noexpand\noexpand\FirstLineOfTheorem}{\hfill$\Box$\end{#2}}}

\MakeFact{Definition}{definition}
\MakeFact{Corollary}{corollary}
\MakeFact{Lemma}{lemma}
\MakeFact{Theorem}{theorem}

\newcommand\Section[1]{\section{#1}}
\newcommand\Subsection[1]{\subsection{#1}}
\newcommand\Subsubsection[1]{\subsubsection{#1}}

\newcommand\Footnote[2][\empty] {\ifx#1\empty\kern-0.1em\else\if*#1\relax\else\kern-#1\fi\fi\footnote{#2}} 

\newcommand\Cor[1] {Cor.~\ref{#1}}

\newcommand\Def[1] {Def.~\ref{#1}}
\newcommand\Eqn[1] {(\ref{#1})}
\newcommand\Lem[1] {Lem.~\ref{#1}}
\newcommand\Thm[1] {Thm.~\ref{#1}}
\newcommand\Sec[1] {\S\ref{#1}}
\newcommand\App[1] {App.~\ref{#1}}
\newcommand\Fig[2][\empty] {\ifx#1\empty Fig.~\ref{#2}\else Fig.~\ref{#2}(#1)\fi}

\newcommand\Itm[1] {(\ref{#1})}

\newenvironment{Proof}{\par\smallskip\noindent\textbf{Proof}\quad}{}

\lstnewenvironment{Prog} {
	\lstset{
		language=csh,
		basicstyle=\ttfamily,	
  		breaklines=false,	
  		showspaces=false,
  		keywords={vis,hid,var,while,if,end,begin,to,then,fi,leak},
  		columns=flexible,
  		escapechar=\%,	
  		mathescape=true
	}
}{}

\lstnewenvironment{ASCII} {
	\lstset{
		language=csh,
		basicstyle=\ttfamily,
  		showspaces=false,
  		breaklines=true,	
  		breakindent=0pt,
  		keywords={},
  		columns=flexible,
  		escapechar=\%,	
  		mathescape=true
	}
}{}

\newenvironment{Reason}{\begin{tabbing}\hspace{4em}\= \hspace{1cm} \= \kill}
    {\end{tabbing}\vspace{-1em}}
\newcommand\Step[2] {#1 \> $\begin{array}[t]{@{}llll}#2\end{array}$ \\}
\newcommand\StepR[3] {#1 \> $\begin{array}[t]{@{}llll}#3\end{array}$
    \` {\RF \makebox[0pt][r]{\begin{tabular}[t]{r}``#2''\end{tabular}}} \\}
\newcommand\WideStepR[3] {#1 \>
    $\begin{array}[t]{@{}ll}~\\#3\end{array}$ \`
    {\RF \makebox[0pt][r]{\begin{tabular}[t]{r}``#2''\end{tabular}}} \\}
\newcommand\Space {~ \\}
\newcommand\RF {\small}

%% file: Macros.tex
\newcommand\Abs[1]		{|#1|}
\renewcommand\AC			{\textit{AC}} 
\newcommand\ACT			{\textit{ACT}}
\newcommand\Comp		{\mathbin{\circ}} 
\newcommand\Defs			{{:=}\,}
\newcommand\Dist			{{\mathbb D}}
\newcommand\Exp			{{\cal E}}
\newcommand\False			{\textit{false}}
\newcommand\Fun			{\mathbin{\rightarrow}} 
\newcommand\In			{{:}\,} 
\newcommand\Lgg			{\mathop{\textrm{lgg}}}
\newcommand\Max			{\mathbin{\textsf{max}}}
\newcommand\Min			{\mathbin\textsf{min}}
\newcommand\NF[2]			{\nicefrac{#1}{#2}}
\newcommand\NNReal		{{\mathbb R}^\geq}
\newcommand\PC[1]			{\mathbin{{}_{#1}\kern-.05em\oplus}}
\newcommand\PCF			{\PC{\NF{1}{2}}}
\newcommand\pGCL		{\textit{pGCL}}
\newcommand\Pow			{{\mathbb P}}
\newcommand\Progress		{\emph{progress}}
\DeclareMathSymbol\RRightarrow{\mathrel}{AMSa}{"56}
\newcommand\SMart 		{\emph{super-martingale}}
\newcommand\True			{\textit{true}}
\newcommand\Wide[1] 		{~~~#1~~~} 
\newcommand\WideRm[1]	{~~~\textrm{#1}~~~} 
\newcommand\WLP			{\textsf{wlp}}
\newcommand\WP			{\textsf{wp}}

%% file: MacrosMarkup.tex
\usepackage{amssymb}			
\usepackage[normalem]{ulem}		
\usepackage{bbding}				
\usepackage{endnotes}			
\usepackage{hyperref}			

\definecolor{PurplePlum}{rgb}{0.1,0,0.55} 
\definecolor{Brown}{rgb}{0.5,.25,0}
\definecolor{Orange}{rgb}{1,.6,0}
\definecolor{Gray}{rgb}{.7,.7,.7}
\definecolor{DarkGreen}{rgb}{0,.6,0}

\newif\ifBleck
\newcommand\Bleck {\Blecktrue} 

\newcommand\Colour[1] {\color{#1}}

\newcommand\ToCLinks {
 \ifBleck\else 
   \expandafter\def\csname@oddfoot\endcsname{
    {\Colour{blue}\mbox{\hyperlink{w1619}{\sf$\rightarrow$~top}\quad
    				    \hyperlink{w1031}{\sf$\rightarrow$~toc}\quad
                                      \hyperlink{GreenRoom}{\sf$\rightarrow$~gr}\quad
                                      \hyperlink{EndNotes}{\sf$\rightarrow$~en}\quad
                                      \hyperlink{Sargasso}{\sf$\rightarrow$~sg}}\hfill
   }}
   \expandafter\def\csname@evenfoot\endcsname{
    {\Colour{blue}\mbox{\hyperlink{w1619}{\sf$\rightarrow$~top}\quad
                                      \hyperlink{w1031}{\sf$\rightarrow$~toc}\quad
                                      \hyperlink{GreenRoom}{\sf$\rightarrow$~gr}\quad
                                      \hyperlink{EndNotes}{\sf$\rightarrow$~en}\quad
                                      \hyperlink{Sargasso}{\sf$\rightarrow$~sg}}\hfill
   }}
 \fi
}

\newif\ifEndNotes 

\newcommand\FnSym{{\scriptsize\PencilLeftDown\kern.1em}} 
\newcommand\EnSym {{$\bigtriangledown$}}
\newcommand\Fn[1] {[\FnSym\ref{#1}]} 

\newif\ifMakeMarkupsCalled 
\newif\ifSuppress 
\newcommand\MakeMarkups[3][.]{
 \Suppressfalse 
 \ifBleck\Suppresstrue\fi
 \ifx0#1\Suppresstrue\fi
 \ifx1#1\Suppressfalse\fi
 \expandafter\providecommand\csname#2x\endcsname {} 
 \ifSuppress\expandafter\renewcommand\csname#2x\endcsname{\relax}\else
                   \expandafter\renewcommand\csname#2x\endcsname{#3}\fi 
 \expandafter\providecommand\csname#2\endcsname {} 
 \ifSuppress\expandafter\renewcommand\csname#2\endcsname[1]{##1}\else
                   \expandafter\renewcommand\csname#2\endcsname[1]{{\csname#2x\endcsname##1}}\fi 
 \expandafter\providecommand\csname#2d\endcsname {} 
 \ifSuppress\expandafter\renewcommand\csname#2d\endcsname[1]{\relax}\else
                   \expandafter\renewcommand\csname#2d\endcsname[1]{{\csname#2x\endcsname\sout{##1}}}\fi 
 \expandafter\providecommand\csname#2r\endcsname {} 
 \ifSuppress\expandafter\renewcommand\csname#2r\endcsname[2]{{##2}}\else
                   \expandafter\renewcommand\csname#2r\endcsname[2]{\csname#2d\endcsname{##1} \csname#2\endcsname{##2}}\fi 
 \expandafter\providecommand\csname#2t\endcsname {} 
 \ifSuppress\expandafter\renewcommand\csname#2t\endcsname[1]{\relax}\else
                   \expandafter\renewcommand\csname#2t\endcsname[1]{{\csname#2x\endcsname{$\langle\!\langle$##1$\rangle\!\rangle$}}}\fi 
 \expandafter\providecommand\csname#2b\endcsname {} 
 \ifSuppress\expandafter\renewcommand\csname#2b\endcsname[1][\empty]{\relax}\else 
                   \expandafter\renewcommand\csname#2b\endcsname[1][\empty]{\ifx##1\empty
                   	\label{#2-bookmark} 
                          \marginpar [\raggedleft\csname#2\endcsname{{\footnotesize\fbox{#2 working here}}~$\Longrightarrow$}]
                                            {\csname#2\endcsname{$\Longleftarrow$~{\footnotesize\fbox{#2 working here}}}}
                   \else 
                   	\marginpar [\raggedleft\csname#2\endcsname{\ifx##1\empty\empty\else\fbox{\tiny\parbox{8em}{\raggedright##1}}~\fi$\Longrightarrow$}]
                                            {\csname#2\endcsname{$\Longleftarrow$\ifx##1\empty\empty\else~{\tiny\fbox{\parbox{8em}{\raggedright##1}}}\fi}}\fi}\fi
 \expandafter\providecommand\csname#2TD\endcsname {} 
 \ifSuppress\expandafter\renewcommand\csname#2TD\endcsname{\relax}\else
                   \expandafter\renewcommand\csname#2TD\endcsname{\csname#2\endcsname{\fbox{\texttt{#2} to do}}}\fi 
 \expandafter\providecommand\csname#2Bar\endcsname {} 
 \ifSuppress\expandafter\renewcommand\csname#2Bar\endcsname{\relax}\else
                   \expandafter\renewcommand\csname#2Bar\endcsname{\csname#2\endcsname{\scriptsize\XSolidBrush}}\fi 
 \expandafter\providecommand\csname#2f\endcsname {} 
 \ifSuppress\expandafter\renewcommand\csname#2f\endcsname[2][]{\relax}\else
  \expandafter\renewcommand\csname#2f\endcsname[2][\empty]{ 
    {\mbox{\csname#2x\endcsname\tiny$\boxtimes$}\marginpar{\csname#2x\endcsname\fbox{\FnSym\footnotemark}}\relax 
    \footnotetext{\csname#2x\endcsname##2}}}\fi
 \expandafter\providecommand\csname#2e\endcsname {} 
 \ifSuppress\expandafter\renewcommand\csname#2e\endcsname[1]{\relax}\else
  \expandafter\renewcommand\csname#2e\endcsname[1]{
   \EndNotestrue 
   \mbox{\scriptsize\csname#2x\endcsname$\boxtimes$}\relax
   \marginpar{\csname#2x\endcsname\fbox{\EnSym\endnotemark
                      \hypertarget{ENmark\thepage-\theendnote}{}~\hyperlink{ENtext\thepage-\theendnote}{{\Colour{blue}$\downarrow$}}}
   }
   { 
    \def\zz{\noexpand#3}
    \edef\z{~{\zz[Endnote \theendnote\ on p.\noexpand\hypertarget{ENtext\thepage-\theendnote}{}\thepage
                ~\noexpand\hyperlink{ENmark\thepage-\theendnote}{{\noexpand\Colour{blue}$\uparrow$}}]}
    }
    \expandafter\endnotetext\expandafter{\z\vspace{2ex}\\ ##1\newpage}
   } 
  }\fi
 \expandafter\providecommand\csname#2fe\endcsname {} 
 \ifSuppress\expandafter\renewcommand\csname#2fe\endcsname[2][]{\relax}\else 
  \expandafter\renewcommand\csname#2fe\endcsname[2][]{ 
   \def\File{##1}\relax
   \ifx\File\empty\csname#2f\endcsname{##2}\else 
    \EndNotestrue 
    \mbox{\scriptsize\csname#2x\endcsname$\boxtimes$}
    \marginpar{\csname#2x\endcsname\fbox{\FnSym\footnotemark}}\relax
    \footnotetext{~\csname#2x\endcsname##2\
                         --- See [\EnSym\endnotemark\hypertarget{ENmark\thepage-\theendnote}{}
                         \kern-.2em\hyperlink{ENtext\thepage-\theendnote}{{\Colour{blue}$\downarrow$}}].}\relax
   { 
     \def\zz{\noexpand#3}
     \edef\z{~{\zz[Footnote~\thefootnote~on~p.\noexpand\hypertarget{ENtext\thepage-\theendnote}{}\thepage
                 ~\noexpand\hyperlink{ENmark\thepage-\theendnote}
                 {{\noexpand\Colour{blue}\kern-0.1em$\uparrow$}]}}
                 {\noexpand\footnotesize\noexpand\newline\noexpand\hspace*{2em} (~from file {\noexpand\tt\File.tex}~)}
     }    
     \expandafter\endnotetext\expandafter{\z~\par\input{##1}\newpage}
    } 
   \fi 
  } 
 \fi 
 \ifSuppress\relax\else
  \csname#2f\endcsname{
   $\backslash$\texttt{#2}$\cdots$\ markups are in \textbf{this} colour\ifx#1..\else\ifx1#1.\else, for #1.\fi\fi
   \ifMakeMarkupsCalled\relax\else 
    \begin{quote}\begin{tabular}{l@{\hspace{2em}}p{.8\linewidth}}
     \multicolumn{2}{l}{\texttt{$\backslash$MakeMarkups\ifx#1.\relax\else[#1]\fi\{#2\}\{{\it$\langle$colour command\/$\rangle$}\}}
     				 --- Defines the macros below:}\\
         & see comments at \texttt{$\backslash$MakeMarkups} definition. \\[1ex]
     \texttt{$\backslash$#2\{$\langle$text$\rangle$\}} & Sets \texttt{$\langle$text$\rangle$} in \texttt{#2}'s colour. \\
     \texttt{$\backslash$#2x} & Changes to \texttt{#2}'s colour (until end of context). \\
     \texttt{$\backslash$#2d\{$\langle$text$\rangle$\}} & Sets \texttt{$\langle$text$\rangle$} in \texttt{#2}'s colour with a strikethrough (i.e.\ delete). \\
     \texttt{$\backslash$#2r\{$\langle$this$\rangle$\}\{$\langle$that$\rangle$\}} &
      Strikes through \texttt{$\langle$this$\rangle$} and inserts \texttt{$\langle$that$\rangle$} (i.e.\ replace). \\
     \texttt{$\backslash$#2f\{$\langle$text$\rangle$\}} & Meta-comment: puts \texttt{$\langle$text$\rangle$} in a \texttt{#2}-footnote with a {\tiny$\boxtimes$} in the main text. \\
     \texttt{$\backslash$#2t\{$\langle$text$\rangle$\}} & Use for meta when  \texttt{$\backslash$#2f} isn't allowed (``Not in outer-par mode.'') \\
     \texttt{$\backslash$#2b[$\langle$optional$\rangle$]} & Marginal pointer, with label for hyper-linking directly there. \\
     \texttt{$\backslash$#2e\{$\langle$text$\rangle$\}} & Puts \texttt{$\langle$text$\rangle$} in a \texttt{#2}-endnote with a (big) $\boxtimes$ in the main text. \\[.5ex]
     \texttt{$\backslash$#2fe[$\langle$this$\rangle$]\{$\langle$that$\rangle$\}} & Makes a  \texttt{$\backslash$#2f\{$\langle$that$\rangle$\}} that refers to a \\
       & \texttt{$\backslash$#2e\{$\langle$contents of file this.tex$\rangle$\}}. \\ 
       & Without the optional argument, acts as \texttt{$\backslash$#2f\{$\langle$that$\rangle$\}}. \\[.5ex]
     \texttt{$\backslash$#2Bar} & Inserts ``burn after reading'' symbol \csname#2Bar\endcsname, meaning
      \begin{quote}\begin{itemize}\setlength\itemsep{0pt}
       \item If yours is the only \csname#2Bar\endcsname\ in this (presumably someone else's) footnote, and you are happy that the footnote has been addressed,
       go ahead and comment-out the whole footnote. (The \csname#2Bar\endcsname\ is their request for you to ``approve and remove''.)
       \item If you are not happy, delete only your \csname#2Bar\endcsname\ and follow-on in the footnote
        (in your colour, i.e.\ with \texttt{$\backslash$#2x}) saying why you are not happy.
       \item If you are happy, but there are others' burn-after-reading symbols as well as yours, just delete yours; the other people have not yet responded.
      \end{itemize}
      \end{quote}
      The idea is that when everyone's happy, the last person will comment-out the meta-text. \\[0.5ex]
     \texttt{$\backslash$#2TD} & Inserts {\csname#2TD\endcsname}\ . \\
    \end{tabular}\end{quote}
   \fi
  }
  \MakeMarkupsCalledtrue 
 \fi
}

\newcommand\MakeGreenRoom {\ifBleck\relax\else
 \hrule
 ~\\\begin{center}\Huge \hypertarget{GreenRoom}{Green Room}
 \end{center}~\\
 \hrule
\fi}

\newcommand\MakeEndNotes {\ifBleck\relax\else
 \hrule~\\\begin{center}\ifEndNotes\Huge Endnotes\else\textit{No endnote macros were called in this run.}\fi\end{center}~\\\hrule
 {\parindent 0pt \parskip 2ex \def\enotesize{\normalsize} \def\notesname{} 
 \ifEndNotes\theendnotes\fi} 
\fi}

\newcommand\MakeSargasso {
 \hypertarget{Sargasso}{}
 \newcommand\NewLabel[1] {\OldLabel{Sargasso-##1}} 
 \newcommand\NewRef[1] 
 {\expandafter\ifx\csname r@Sargasso-##1\endcsname\relax\OldRef{##1}\else\OldRef{Sargasso-##1}\fi}
 \let\OldLabel\label \let\label\NewLabel
 \let\OldRef\ref \let\ref\NewRef
 \ifBleck\end{document}\else
  \hrule
  ~\\\begin{center}\Huge Sargasso
  \end{center}~\\
  \hrule
 \fi
}

\newcommand\Cite[2][\empty] {{\color{red}\ifx#1\empty[#2]\else[#2,~#1]\fi}}


%% file: MacrosLabels.tex

%

\newif\ifShowLabels
\newcommand\SL[1] {\mbox{\tt\scriptsize[#1]}}
\newcommand\LabelWithShow[1] {\ifShowLabels\ifinner\SL{#1}\else\ifmmode\SL{#1}\else\marginpar{\SL{#1}}\fi\fi\fi\Label{#1}}
\let\Label\label \let\label\LabelWithShow

\newcommand\ShowLabels {\ShowLabelstrue} 
\newcommand\NoShowLabels {\ShowLabelsfalse} 

\newif\ifShowLabelsWas
\newenvironment{Equation}{\ShowLabelsWasfalse\ifShowLabels\ShowLabelsWastrue\begin{equation}\let\LabelWas\Label\let\Label\label\let\label\LabelWithShow\else\begin{equation}\fi}{\end{equation}\ifShowLabelsWas\let\label\Label\let\Label\LabelWas\fi\ShowLabelsWasfalse}